\documentclass[10pt,journal,compsoc]{IEEEtran}
\usepackage{amsmath}
\usepackage{color}
\usepackage{amssymb}
\usepackage{algorithm}
\usepackage{algorithmic}
\usepackage[pagebackref=true,colorlinks,bookmarks=false]{hyperref}
\ifCLASSOPTIONcompsoc
\else
\fi
\ifCLASSINFOpdf
\else
\fi

\newtheorem{definition}{Definition}

\begin{document}
%
\title{Hashing for Similarity Search: A Survey}
%
%
%
%

\author{Jingdong~Wang,
Heng Tao Shen,
Jingkuan Song,
and Jianqiu Ji
\IEEEcompsocitemizethanks{\IEEEcompsocthanksitem
J. Wang is with Microsoft Research,
Beijing, P.R. China. \protect\\
E-mail: jingdw@microsoft.com
\IEEEcompsocthanksitem
J. Song and H.T. Shen are with School of Information Technology and Electrical Engineering,
The University of Queensland, Australia.\protect\\
Email:\{jk.song,shenht\}@itee.uq.edu.au
\IEEEcompsocthanksitem
J. Ji is with Department of Computer Science and Technology, Tsinghua University,
Beijing, P.R. China. \protect \\
E-mail: jijq10@mails.tsinghua.edu.cn
}}

%
%


\markboth{}%
{Shell \MakeLowercase{\textit{et al.}}: Hashing for Approximate Nearest Neighbor Search: A Survey}

%


\IEEEspecialpapernotice{\today}

\IEEEtitleabstractindextext{%
\begin{abstract}
Similarity search (nearest neighbor search)
is a problem
of pursuing the data items whose distances to a query item
are the smallest
from a large database.
Various methods have been developed
to address this problem,
and recently a lot of efforts
have been devoted to approximate search.
In this paper,
we present a survey on one of the main solutions, hashing,
which has been widely studied
since the pioneering work locality sensitive hashing.
We divide the hashing algorithms two main categories:
locality sensitive hashing,
which designs hash functions without exploring the data distribution
and learning to hash,
which learns hash functions according the data distribution,
and review them
from various aspects,
including
hash function design
and distance measure and search scheme in the hash coding space.
\end{abstract}

}

\maketitle

\IEEEdisplaynontitleabstractindextext

%
\IEEEpeerreviewmaketitle

\section{Introduction}
\label{sec:introduction}
The problem of similarity search,
also known as
nearest neighbor search,
proximity search,
or close item search,
is to find an item
that is the nearest
to a query item,
called nearest neighbor,
under some distance measure
from a search (reference) database.
In the case that the reference database
is very large
or that the distance computation
between the query item and the database item
is costly,
it is often computationally infeasible
to find the exact nearest neighbor.
Thus,
a lot of research efforts
have been devoted
to approximate nearest neighbor search
that is shown to be enough and useful
for many practical problems.

Hashing is one of the popular solutions
for approximate nearest neighbor search.
In general, hashing
is an approach of
transforming the data item
to a low-dimensional representation,
or equivalently a short code consisting of a sequence of bits.
The application of hashing to approximate nearest neighbor search
includes two ways:
indexing data items using hash tables
that is formed by storing the items
with the same code in a hash bucket,
and approximating the distance
using the one computed with short codes.

The former way
regards the items
lying the buckets corresponding to the codes
of the query
as the nearest neighbor candidates,
which exploits the locality sensitive property
that similar items have larger probability
to be mapped to the same code
than dissimilar items.
The main research efforts along this direction
consist of designing hash functions satisfying the locality sensitive property
and designing efficient search schemes
using and beyond hash tables.

The latter way
ranks the items according to
the distances computed using the short codes,
which exploits the property
that the distance computation
using the short codes
is efficient.
The main research effort along this direction
is to design the effective ways to compute the short codes
and design the distance measure using the short codes
guaranteeing the computational efficiency
and preserving the similarity.

\section{Overview}
\label{sec:overview}
\subsection{The Nearest Neighbor Search Problem}
\subsubsection{Exact nearest neighbor search}
Nearest neighbor search,
also known as similarity search,
proximity search,
or close item search,
is defined as:
Given a query item $\mathbf{q}$,
the goal is to find
an item $\operatorname{NN}(\mathbf{q})$,
called nearest neighbor,
from a set of items $\mathcal{X} = \{\mathbf{x}_1, \mathbf{x}_2, \cdots, \mathbf{x}_N\}$
so that
$\operatorname{NN}(\mathbf{q}) = \arg\min_{\mathbf{x} \in \mathcal{X}} \operatorname{dist}(\mathbf{q}, \mathbf{x})$,
where $\operatorname{dist}(\mathbf{q}, \mathbf{x})$ is a distance
computed between $\mathbf{q}$ and $\mathbf{x}$.
A straightforward generalization
is a $K$-NN search,
where $K$-nearest neighbors ($\operatorname{KNN}(\mathbf{q})$) are needed to be found.

The problem is not fully specified without
the distance
between an arbitrary pair of items $\mathbf{x}$ and $\mathbf{q}$.
As a typical example,
the search (reference) database $\mathcal{X}$ lies in
a $d$-dimensional space $\mathbb{R}^d$
and the distance is induced by an $l_s$ norm,
$\|\mathbf{x} - \mathbf{q}\|_s = (\sum_{i=1}^d |x_i - q_i|^s)^{1/s}$.
The search problem under the Euclidean distance, i.e., the $l_2$ norm,
is widely studied.
Other notions of search database,
such as each item formed by a set,
and distance measure,
such as $\ell_1$ distance,
cosine similarity and so on
are also possible.

The fixed-radius near neighbor ($R$-near neighbor) problem,
an alternative of nearest neighbor search,
is defined as:
Given a query item $\mathbf{q}$,
the goal is to find
the items $\mathcal{R}$
that are within the distance $C$ of $\mathbf{q}$,
$\mathcal{R} =
\{\mathbf{x} | \operatorname{dist}(\mathbf{q}, \mathbf{x}) \leqslant R, \mathbf{x} \in \mathcal{X}\}$.


\subsubsection{Approximate nearest neighbor search}
There exists efficient algorithms
for exact nearest neighbor and $R$-near neighbor search problems
in low-dimensional cases.
It turns out that the problems become
hard in the large scale high-dimensional case
and even most algorithms take higher computational cost
than the naive solution, linear scan.
Therefore,
a lot of recent efforts
are moved
to approximate nearest neighbor search problems.
The $(1+\epsilon)$-approximate nearest neighbor search problem,
$\epsilon > 0$,
is defined as:
Given a query $\mathbf{x}$,
the goal is to
find an item $\mathbf{x}$
so that $\operatorname{dist}(\mathbf{q}, \mathbf{x})
\leqslant (1+\epsilon) \operatorname{dist}(\mathbf{q}, \mathbf{x})^*$,
where $\mathbf{x}^*$ is the true nearest neighbor.
The $c$-approximate $R$-near neighbor search problem
is defined as:
Given a query $\mathbf{x}$,
the goal is to
find some item $\mathbf{x}$,
called $cR$-near neighbor,
so that $\operatorname{dist}(\mathbf{q}, \mathbf{x})
\leqslant cR$,
where $\mathbf{x}^*$ is the true nearest neighbor.

\subsubsection{Randomized nearest neighbor search}
The randomized search problem aims to
report the (approximate) nearest (or near) neighbors
with probability
instead of deterministically.
There are two widely-studied randomized search problems:
randomized $c$-approximate $R$-near neighbor search
and
randomized $R$-near neighbor search.
The former one is defined as:
Given a query $\mathbf{x}$,
the goal is to report some $cR$-near neighbor of the query $\mathbf{q}$
with probability $1-\delta$,
where $0<\delta<1$.
The latter one is defined as:
Given a query $\mathbf{x}$,
the goal is to report some $R$-near neighbor of the query $\mathbf{q}$
with probability $1-\delta$.

\subsection{The Hashing Approach}
The hashing approach aims to
map the reference and/or query items
to the target items
so that approximate nearest neighbor search
can be efficiently and accurately performed
using the target items and possibly
a small subset of the raw reference items.
The target items are
called hash codes
(also known as hash values,
simply hashes).
In this paper, we may also call it
short/compact code interchangeably.

Formally,
the hash function is defined as:
$y = h(\mathbf{x})$,
where $y$ is the hash code
and $h(\cdot)$ is the function.
In the application
to approximate nearest neighbor search,
usually several hash functions are used
together
to compute the hash code:
$\mathbf{y} = \mathbf{h}(\mathbf{x})$,
where $\mathbf{y} = [y_1~y_2~\cdots~y_M]^T$
and $[h_1(\mathbf{x})~h_2(\mathbf{x})~\cdots~h_M(\mathbf{x})]^T$.
Here we use a vector $\mathbf{y}$
to represent the hash code
for presentation convenience.

There are two basic strategies for
using hash codes to perform
nearest (near) neighbor search:
hash table lookup
and Fast distance approximation.

\subsubsection{Hash table lookup.}
The hash table is a data structure
that is composed of buckets,
each of which is indexed by a hash code.
Each reference item $\mathbf{x}$ is placed
into a bucket $h(\mathbf{x})$.
Different from the conventional hashing algorithm
in computer science
that avoids collisions
(i.e., avoids mapping two items
into some same bucket),
the hashing approach using a hash table
aims to maximize the probability
of collision of near items.
Given the query $\mathbf{q}$,
the items lying in the bucket $h(\mathbf{q})$
are retrieved as near items of $\mathbf{q}$.

To improve the recall,
$L$ hash tables are constructed,
and the items lying in the $L$ ($L'$, $L'< L$) hash buckets
$h_1(\mathbf{q}), \cdots, h_L(\mathbf{q})$
are retrieved as near items of $\mathbf{q}$
for randomized $R$-near neighbor search
(or randomized $c$-approximate $R$-near neighbor search).
To guarantee the precision,
each of the $L$ hash codes, $\mathbf{y}_i$,
needs to be a long code,
which means that
the total number of the buckets is too large
to index directly.
Thus,
only the nonempty buckets
are retained
by resorting to convectional hashing of the hash codes
$h_l(\mathbf{x})$.

\subsubsection{Fast distance approximation.}
The direct way is to perform an exhaustive search:
compare the query with each reference item
by fast computing the distance between the query and the hash code of the reference item
and retrieve the reference items with the smallest distances
as the candidates of nearest neighbors,
which is usually followed by
a reranking step:
rerank the nearest neighbor candidates
retrieved with hash codes
according to the true distances
computed using the original features
and attain the $K$ nearest neighbors
or $R$-near neighbor.

This strategy exploits two advantages
of hash codes.
The first one is that the distance using hash codes
can be efficiently computed
and the cost is much smaller
than that of the computation
in the input space.
The second one is that
the size of the hash codes
is much smaller than the input features
and hence can be loaded into memory,
resulting the disk I/O cost reduction
in the case the original features are too large
to be loaded into memory.

One practical way of speeding up the search
is to perform a non-exhaustive search:
first retrieve a set of candidates using inverted index
and then compute the distances of the query with the candidates using the short codes.
Other research efforts includes organizing the hash codes with a data structure,
such as a tree or a graph structure,
to avoid exhaustive search.


\subsection{Organization of This Paper}
The organization of the remaining part is given as follows.
Section~\ref{sec:LSH1} presents the definition of the locality sensitive hashing (LSH) family
and the instances of LSH with various distances.
Section~\ref{sec:LSH2} presents some research works
on how to perform efficient search given LSH codes
and model and analyze LSH in aspects
Sections~\ref{sec:LTH1},~\ref{sec:LTH2},and ~\ref{sec:LTH3}
review the learning-to-hash algorithms.
Finally,
Section~\ref{sec:con} concludes this survey.

\section{Locality Sensitive Hashing: Definition and Instances}
\label{sec:LSH1}
The term ``locality-sensitive hashing'' (LSH) was introduced in 1998~\cite{IndykM98}, to name a randomized hashing framework for efficient approximate nearest neighbor (ANN) search in high dimensional space. It is based on the definition of LSH family $\mathcal{H}$, a family of hash functions mapping similar input items to the same hash code with higher probability than dissimilar items. However, the first specific LSH family, min-hash, was invented in 1997 by Andrei Broder~\cite{Broder97}, for near-duplicate web page detection and clustering, and it is one of the most popular LSH method that is extensively-studied in theory and widely-used in practice.

Locality-sensitive hashing was first studied by the theoretical computer science community. The theoretical research mainly focuses on three aspects. The first one is on developing different LSH families for various distances or similarities, for example, p-stable distribution LSH for $\ell_p$ distance \cite{DatarIIM04}, sign-random-projection (or sim-hash) for angle-based distance \cite{Charikar02}, min-hash for Jaccard coefficient \cite{Broder97,BroderGMZ97} and so on, and many variants are developed based on these basic LSH families \cite{DasguptaKS11}. The second one is on exploring the theoretical boundary of the LSH framework, including the bound on the search efficiency (both time and space) that the best possible LSH family can achieve for certain distances and similarities \cite{DatarIIM04,MotwaniNP07,ODonnellWZ11}, the tight characteristics for a similarity measure to admit an LSH family \cite{Charikar02,ChierichettiK12a}, and so on.
The third one focuses on improving the search scheme of the LSH methods, to achieve theoretically provable better search efficiency \cite{Panigrahy06,DasguptaKS11}.

Shortly after it was proposed by the theoretical computer science community, the database and related communities began to study LSH, aiming at building real database systems for high dimensional similarity search. Research from this side mainly focuses on developing better data structures and search schemes that lead to better search quality and efficiency in practice \cite{LvJWCL07,GanFFN12}. The quality criteria include precision and recall, and the efficiency criteria are commonly the query time, storage requirement, I/O consumption and so on. Some of these work also provide theoretical guarantees on the search quality of their algorithms \cite{GanFFN12}.

In recent years, LSH has attracted extensive attention from other communities including computer vision (CV), machine learning, statistics, natural language processing (NLP) and so on. For example, in computer vision, high dimensional features are often required for various tasks, such as image matching, classification. LSH, as a probabilistic dimension reduction method, has been used in various CV applications which often reduce to approximate nearest neighbor search \cite{ChumPIZ07,ChumPZ08}. However, the performance of LSH is limited due to the fact that it is totally probabilistic and data-independent, and thus it does not take the data distribution into account. On the other hand, as an inspiration of LSH, the concept of ``small code'' or ``compact code'' has become the focus of many researchers from the CV community, and many learning-based hashing methods have come in to being \cite{WeissTF08}\cite{WangKC10a}\cite{WangKC10b}\cite{GongL11}\cite{LiuWKC11}\cite{XuWLZLY11}\cite{WangKC12}\cite{GongKVL12}\cite{LiuWJJC12}\cite{WangWYL13}\cite{GongLGP13}. These methods aim at learning the hash functions for better fitting the data distribution and labeling information, and thus overcoming the drawback of LSH. This part of the research often takes LSH as the baseline for comparison.

The machine learning and statistics community also contribute to the study of LSH. Research from this side often view LSH as a probabilistic similarity-preserving dimensionality reduction method, from which the hash codes that are produced can provide estimations to some pairwise distance or similarity. This part of the study mainly focuses on developing variants of LSH functions that provide an (unbiased) estimator of certain distance or similarity, with smaller variance \cite{LiCH06,JiLYZT12,LiOZ12,JiLYTZ13}, or smaller storage requirement of the hash codes \cite{LiK10b,LiKG10a}, or faster computation of hash functions \cite{LiHC06,LiOZ12,JiLYTZ13,ShrivastavaL14}. Besides, the machine learning community also devotes to developing learning-based hashing methods.

In practice, LSH is widely and successfully used in the IT industry, for near-duplicate web page and image detection, clustering and so on. Specifically, The Altavista search engine uses min-hash to detect near-duplicate web pages \cite{Broder97,BroderGMZ97}, while Google uses sim-hash to fulfill the same goal \cite{MankuJS07}.

In the subsequent sections,
we will first introduce different LSH families for various kinds of distances or similarities,
and then we review the study focusing on the search scheme and the work devoted to modeling LSH and ANN problem.

\subsection{The Family}
The locality-sensitive hashing (LSH) algorithm is introduced in~\cite{IndykM98, GionisIM99}, to solve the $(R, c)$-near neighbor problem.
It is based on the definition of LSH family $\mathcal{H}$,
a family of hash functions mapping similar input items to the same hash code with higher probability than dissimilar items.
Formally, an LSH family is defined as follows:

\label{def:knn}
\begin{definition}[Locality-sensitive hashing]
A family of $\mathcal{H}$
is called $(R, cR, P_1, P_2)$-sensitive
if for any two items $\mathbf{p}$ and $\mathbf{q}$,
\begin{itemize}
  \item if $\operatorname{dist}(\mathbf{p}, \mathbf{q}) \leqslant R$, then $\operatorname{Prob}[h(\mathbf{p}) = h(\mathbf{q})] \geqslant P_1$,
  \item if $\operatorname{dist}(\mathbf{p}, \mathbf{q}) \geqslant cR$, then $\operatorname{Prob}[h(\mathbf{p}) = h(\mathbf{q})] \leqslant P_2$.
\end{itemize}
\end{definition}
Here $c > 1$, and $P_1 > P_2$.
The parameter $\rho = \frac{\log(1/P_1)}{\log(1/P_2)}$
governs the search performance,
the smaller $\rho$,
the better search performance.
Given such an LSH family for distance measure $\operatorname{dist}$,
there exists an algorithm for $(R,c)$-near neighbor problem which uses $O(dn+n^{1+\rho})$ space,
with query time dominated by $O(n^{\rho})$ distance computations and $O(n^{\rho}\log_{1/p_2}n)$ evaluations of hash functions \cite{DatarIIM04}.

The LSH scheme indexes all items in hash tables and searches for near items via hash table lookup.
The hash table is a data structure
that is composed of buckets,
each of which is indexed by a hash code.
Each reference item $\mathbf{x}$ is placed
into a bucket $h(\mathbf{x})$.
Different from the conventional hashing algorithm
in computer science
that avoids collisions
(i.e., avoids mapping two items
into some same bucket),
the LSH approach
aims to maximize the probability
of collision of near items.
Given the query $\mathbf{q}$,
the items lying in the bucket $h(\mathbf{q})$
are considered as near items of $h(\mathbf{q})$.

Given an LSH family $\mathcal{H}$,
the LSH scheme amplifies the gap
between the high probability $P_1$
and the low probability $P_2$
by concatenating several functions.
In particular,
for parameter $K$,
$K$ functions
$h_{1}(\mathbf{x}),...,h_{K}(\mathbf{x})$,
where $h_k$
($1\leqslant k \leqslant K$)
are chosen independently and uniformly
at random from $\mathcal{H}$, form a compound hash function $g(\mathbf{x}) = (h_{1}(\mathbf{x}),\cdots, h_{K}(\mathbf{x}))$. The output of this compound hash function identifies a bucket id in a hash table. However, the concatenation of $K$ functions also reduces the chance of collision between similar items. To improve the recall,
$L$ such compound hash functions $g_{1},g_{2},...,g_{L}$ are sampled independently, each of which corresponds to a hash table.
These functions are used to hash each data point
into $L$ hash codes, and $L$ hash tables are constructed to index the buckets corresponding to these hash codes respectively. The items lying in the $L$ hash buckets
are retrieved as near items of $h(\mathbf{q})$
for randomized $R$-near neighbor search
(or randomized $c$-approximate $R$-near neighbor search).

In practice, to guarantee the precision,
each of the $L$ hash codes, $g_l(\mathbf{x})$,
needs to be a long code (or $K$ is large),
and thus the total number of the buckets is too large
to index directly.
Therefore, only the nonempty buckets
are retained
by resorting to conventional hashing of the hash codes
$g_l(\mathbf{x})$.

There are different kinds of LSH families for different distances or similarities, including $\ell_p$ distance, $arccos$ or angular distance, Hamming distance, Jaccard coefficient and so on.

\subsection{$\ell_p$ Distance}
\subsubsection{LSH with $p$-stable distributions}
The LSH scheme based on the $p$-stable distributions,
presented in~\cite{DatarIIM04},
is designed to solve the search problem under the $\ell_p$ distance
$\|\mathbf{x}_i - \mathbf{x}_j\|_p$, where $p\in(0,2]$.
The $p$-stable distribution is defined as:
A distribution $\mathcal{D}$ is called $p$\emph{-stable},
where $p \geqslant 0$,
if
for any $n$ real numbers $v_1 \cdots v_n$
and i.i.d. variables $X_1 \cdots X_n$
with distribution $\mathcal{D}$,
the random variable $\sum_{i=1}^nv_iX_i$
has the same distribution as the variable
$(\sum_{i=1}^{n}|v_i|^p)^{1/p}X$,
where $X$ is a random variable with distribution $\mathcal{D}$.
The well-known Gaussian distribution $\mathcal{D}_G$,
defined
by the density function $g(x) = \frac{1}{\sqrt{2\pi}}e^{-x^2/2}$,
is $2$-stable.

In the case that $p=1$,
the exponent $\rho$ is equal to $\frac{1}{c} + O(R/r)$,
and later it is shown in~~\cite{MotwaniNP07} that it is impossible
to achieve $\rho \leqslant \frac{1}{2c}$.
Recent study in~\cite{ODonnellWZ11}
provides more lower bound analysis
for Hamming distance, Euclidean distance,
and Jaccard distance.

The LSH scheme using the $p$-stable distribution to generate hash codes
is described as follows.
The hash function is formulated as
$h_{\mathbf{w}, b}(\mathbf{x}) = \lfloor \frac{\mathbf{w}^T\mathbf{x} + b}{r}  \rfloor$.
Here, $\mathbf{w}$ is a $d$-dimensional vector
with entries chosen independently from a $p$-stable distribution.
$b$ is a real number chosen uniformly from
the range $[0, r]$.
$r$ is the window size, thus a positive real number.

The following equation
can be proved
\begin{align}
P(h_{\mathbf{w}, b} (\mathbf{x}_1) = h_{\mathbf{w}, b} (\mathbf{x}_2))
\int_{0}^r \frac{1}{c}
f_p(\frac{t}{c})(1-\frac{t}{r}) dt,
\end{align}
where $c = \|\mathbf{x}_1 - \mathbf{x}_2\|_p$,
which means that such a hash function belongs to the LSH family
under the $\ell_p$ distance.

Specifically,
to solve the search problem
under the Euclidean distance, the $2$-stable distribution, i.e.,
the Gaussian distribution, is chosen
to generate the random projection $\mathbf{w}$.
In this case ($p=2$),
the exponent $\rho$ drops strictly below
$1/c$
for some (carefully chosen)
finite value of $r$.

It is claimed
that uniform quantization~\cite{LiMS14} without the offset $b$,
$h_{\mathbf{w}}(\mathbf{x}) = \lfloor \frac{\mathbf{w}^T\mathbf{x}}{r}  \rfloor$
is more accurate and uses fewer bits than the scheme with the offset.



\subsubsection{Leech lattice LSH}
Leech lattice LSH~\cite{AndoniI06} is an LSH algorithm
for the search in the Euclidean space.
It is a multi-dimensional version
of the aforementioned approach.
The approach firstly
randomly projects the data points
into $\mathbb{R}^t$,
$t$ is a small super-constant
($=1$ in the aforementioned approach).
The space $\mathbb{R}^t$
is partitioned into cells,
using Leech lattice,
which is a constellation in $24$ dimensions.
The nearest point in Leech lattice can
be found
using a (bounded) decoder which performs only
$519$ floating point operations per decoded point.
On the other hand,
the exponent $\rho(c)$ is quite attractive:
$\rho(2)$ is less than $0.37$.
$E_8$ lattice is used because
its decoding is much cheaper
than Leech lattice
(its quantization performance
is slightly worse)
A comparison of LSH methods for the Euclidean distance is given in~\cite{PauleveJA10}.

\subsubsection{Spherical LSH}
Spherical LSH~\cite{TerasawaT07} is an LSH algorithm
designed for points
that are on a unit hypersphere
in the Euclidean space.
The idea is to
consider the regular polytope,
simplex, orthoplex,
and hypercube, for example,
that are inscribed into the hypersphere
and rotated at random.
The hash function maps a vector
on the hypersphere
into the closest polytope vertex lying on the hypersphere.
It means that
the buckets of the hash function
are the Voronoi cells
of the polytope vertices.
Though there is no theoretic analysis about exponent $\rho$,
the Monte Carlo simulation shows that it is an improvement
over the Leech lattice approach~\cite{AndoniI06}.

\subsubsection{Beyond LSH}
Beyond LSH~\cite{AndoniINR14} improves
the ANN search in the Euclidean space,
specifically solving $(c,1)$-ANN.
It consists of two-level hashing structures:
outer hash table
and inner hash table.
The outer hash scheme aims to
partition the data into buckets
with a filtered out process
such that all the pairs of points in the bucket
are not more than a threshold,
and find a $(1+ 1/c)$-approximation
to the minimum enclosing ball for the remaining points.
The inner hash tables
are constructed
by first computing the center
of the ball corresponding to a non-empty bucket in outer hash tables
and partitioning the points belonging to the ball
into a set of over-lapped subsets,
for each of which the differences of the distance of the points to the center
is within $[-1, 1]$
and the distance of the overlapped area to the center
is within $[0, 1]$.
For the subset, an LSH scheme is conducted.
The query process
first locates a bucket from outer hash tables for a query.
If the bucket is empty, the algorithm stops.
If the distance of the query
to the bucket center is not larger than $c$,
then the points in the bucket are output
as the results.
Otherwise,
the process further checks the subsets in the bucket
whose distances to the query lie in a specific range
and then does the LSH query in those subsets.

\subsection{Angle-Based Distance}
\subsubsection{Random projection}
The LSH algorithm based on random projection~\cite{AndoniI08, Charikar02}
is developed
to solve the near neighbor search problem
under the angle between vectors,
$\theta(\mathbf{x}_i, \mathbf{x}_j)
= \arccos\frac{\mathbf{x}_i^T\mathbf{x}_j}{\|\mathbf{x}_i\|_2\|\mathbf{x}_j\|_2}$.
The hash function is formulated
as $h(\mathbf{x}) = \operatorname{sign}(\mathbf{w}^T\mathbf{x})$,
where $\mathbf{w}$ follows the standard Gaussian distribution.
It is easily shown
that $P(h(\mathbf{x}_i) = h(\mathbf{x}_j))
= 1 - \frac{\theta(\mathbf{x}_i, \mathbf{x}_j)}{\pi}$,
where $\theta(\mathbf{x}_i, \mathbf{x}_j)$
is the angle between $\mathbf{x}_i$
and $\mathbf{x}_j$,
thus such a hash function belongs to the LSH family
with the angle-based distance.

\subsubsection{Super-bit LSH}
Super-bit LSH~\cite{JiLYZT12} aims to improve
the above hashing functions for arccos (angular) similarity,
by dividing the random projections into $G$ groups
then orthogonalizing $B$ random projections for each group,
obtaining new $GB$ random projections
and thus $G$ $B$-super bits.
It is shown that the Hamming distance over the super bits
is an unbiased estimation for the angular distance
and the variance is smaller than the above random projection algorithm.

\subsubsection{Kernel LSH}
Kernel LSH~\cite{KulisG09, KulisG12}
aims to build LSH functions
with the angle
defined in the kernel space,
$\theta(\mathbf{x}_i, \mathbf{x}_j)
= \arccos\frac{\phi{(\mathbf{x}_i)}^T\phi{(\mathbf{x}_j)}}{\|\phi{(\mathbf{x}_i)}\|_2\|\phi{(\mathbf{x}_j)}\|_2}$.
The key challenge
is in constructing a projection vector $\mathbf{w}$
from the Gaussian distribution.
Define $\mathbf{z}_t = \frac{1}{t}\sum_{i \in \mathcal{S}_t} \phi(\mathbf{x}_i)$
where $t$ is a natural number,
and $\mathcal{S}$ is a set of $t$ database items chosen i.i.d..
The central limit theorem shows that
for sufficiently large $t$,
the random variables $\tilde{\mathbf{z}}_t = \sqrt{t}\boldsymbol{\Sigma}^{-1/2}(\mathbf{z}_t - \boldsymbol{\mu})$
follows a Normal distribution $N(\mathbf{0}, \mathbf{I})$.
Then the hash function is given as
\begin{equation}
h(\phi(\mathbf{x})) = \left\{ \begin{array}{l l}
     1 & \quad \text{if $\phi(\mathbf{x})\boldsymbol{\Sigma}^{-1/2}\tilde{\mathbf{z}}_t \geqslant 0$}\\
     0 & \quad \text{otherwise.}
   \end{array} \right.
\end{equation}
The covariance matrix $\boldsymbol{\Sigma}$
and the mean $\boldsymbol{\mu}$
are estimated over a set of randomly chosen $p$ database items, using a technique similar to that used in kernel principal component analysis.

Multi-kernel LSH~\cite{WangJHT10, WangHJT12},
uses multiple kernels instead of a single kernel
to form the hash functions
with assigning the same number of bits to each kernel hash function.
A boosted version of multi-kernel LSH
is presented in~\cite{XiaWHJ12},
which adopts the boosting scheme to
automatically assign various number of bits
to each kernel hash function.

\subsubsection{LSH with learnt metric}
Semi-supervised LSH~\cite{JainKDG08, JainKG08, KulisJG09}
first learns a Mahalanobis metric
from the semi-supervised information
and then
form the hash function
according to the pairwise similarity
$\theta(\mathbf{x}_i, \mathbf{x}_j)
= \arccos\frac{\mathbf{x}_i^T\mathbf{A}\mathbf{x}_j}{\|\mathbf{G}\mathbf{x}_i\|_2\|\mathbf{G}\mathbf{x}_j\|_2}$,
where $\mathbf{G}^T\mathbf{G} = \mathbf{A}$
and $\mathbf{A}$ is the learnt metric from the semi-supervised information.
An extension,
distribution aware LSH~\cite{ZhangZZT13},
is proposed,
which, however,
partitions the data along each projection direction
into multiple parts instead of only two parts.

\subsubsection{Concomitant LSH}
Concomitant LSH~\cite{EshghiR08}
is an LSH algorithm
that uses concomitant rank order statistics
to form the hash functions
for cosine similarity.
There are two schemes:
concomitant min hash
and concomitant min $L$-multi-hash.

Concomitant min hash is formulated as follows:
generate $2^K$ random projections
$\{\mathbf{w}_1, \mathbf{w}_2, \cdots,
\mathbf{w}_{2^K}\}$,
each of which is drawn
independently from the standard normal distribution
$\mathcal{N}(\mathbf{0}, \mathbf{I})$.
The hash code is computed
in two steps:
compute the $2^K$ projections
along the $2^K$ projection directions,
and output the index of the projection direction
along which the projection value is the smallest,
formally written by
$h_c(\mathbf{x}) = \arg\min_{k=1}^{2^K} \mathbf{w}_k^T\mathbf{x}$.
It is shown that
the probability
$\operatorname{Prob}[h_c(\mathbf{x}_1) = h_c(\mathbf{x}_2)]$
is a monotonically increasing function with respect to $\frac{\mathbf{x}_1^T\mathbf{x}_2}{\|\mathbf{x}_1\|_2 \|\mathbf{x}_2\|_2}$.

Concomitant min $L$-multi-hash instead generates
$L$ hash codes:
the indices of the projection directions
along which the projection values are the top $L$ smallest.
It can be shown that
the collision probability is similar to that of Concomitant min hash.

Generating a hash code of length $K=20$
means that
it requires $1,048,576$ random projections
and vector multiplications,
which is too high.
To solve this problem,
a cascading scheme is adopted:
e.g.,
generate two concomitant hash functions,
each of which generates a code of length $10$,
and compose them together,
yielding a code of $20$ bits,
which only requires $2 \times 2^{10}$ random projections
and vector multiplications.
There are two schemes proposed in~\cite{EshghiR08}:
cascade concomitant min \& max hash
that composes the two codes
$[\arg\min_{k=1}^{2^K} \mathbf{w}_k^T\mathbf{x}, \arg\max_{k=1}^{2^K} \mathbf{w}_k^T\mathbf{x}]$,
and
cascade concomitant $L^2$  min \& max hash multi-hash
which is formed using the indices
of the top smallest and largest projection values.

\subsubsection{Hyperplane hashing}
The goal of searching nearest neighbors to a query hyperplane is to
retrieve the points from the database $\mathcal{X}$
that are closest to a query hyperplane
whose normal is given by $\mathbf{n} \in \mathbb{R}^d$.
The Euclidean distance of a point $\mathbf{x}$
to a hyperplane with the normal $\mathbf{n}$
is:
\begin{align}
d(P_{\mathbf{n}}, \mathbf{x}) = \|\mathbf{n}^T\mathbf{x}\|.
\end{align}

The hyperplane hashing family~\cite{JainVG10, Vijayanarasimhan0G14},
under the assumption that the hyperplane passes through origin
and the data points and the normal are unit norm
(which indicates that hyperplane hashing corresponds to search
with absolute cosine similarity),
is defined as follows,
\begin{equation}
h(\mathbf{z}) = \left\{ \begin{array}{l l}
     h_{\mathbf{u}, \mathbf{v}}(\mathbf{z}, \mathbf{z}) & \text{if $\mathbf{z}$ is a database vector}\\
     h_{\mathbf{u}, \mathbf{v}}(\mathbf{z}, -\mathbf{z} & \text{if $\mathbf{z}$ is a query hyperplane normal.}
   \end{array} \right.
\end{equation}
Here $h_{\mathbf{u}, \mathbf{v}}(\mathbf{a}, \mathbf{b}) = [h_{\mathbf{u}}(\mathbf{a})~h_{\mathbf{v}}(\mathbf{b})]
= [\operatorname{sign}(\mathbf{u}^T\mathbf{a}) ~ \operatorname{sign}(\mathbf{v}^T\mathbf{b})]$,
where $\mathbf{u}$ and $\mathbf{v}$
are sampled independently from a standard Gaussian distribution.

It is shown that
the above hashing family belongs to LSH:
it is
$(r, r(1+\epsilon), \frac{1}{4} - \frac{1}{\pi^2}r, \frac{1}{4} - \frac{1}{\pi^2}r(1+\epsilon))$-sensitive
for the angle distance $d_{\theta}(\mathbf{x}, \mathbf{n}) = (\theta_{\mathbf{x}, \mathbf{n}} - \frac{\pi}{2})^2$,
where $r, \epsilon > 0$.
The angle distance is equivalent to the distance of a point to the query hyperplane.

The below family,
called XOR $1$-bit hyperplane hashing,
\begin{equation}
h(\mathbf{z}) = \left\{ \begin{array}{l l}
     h_{\mathbf{u}}(\mathbf{z}) \oplus h_{\mathbf{v}}(\mathbf{z}) & \text{if $\mathbf{z}$ is a database vector}\\
     h_{\mathbf{u}}(\mathbf{z}) \oplus h_{\mathbf{v}}(-\mathbf{z}) &\text{if $\mathbf{z}$ is a hyperplane normal,}
   \end{array} \right.
\end{equation}
is shown to be
$(r, r(1+\epsilon), \frac{1}{2} - \frac{1}{\pi^2}r, \frac{1}{2} - \frac{1}{\pi^2}r(1+\epsilon))$-sensitive
for
the angle distance $d_{\theta}(\mathbf{x}, \mathbf{n}) = (\theta_{\mathbf{x}, \mathbf{n}} - \frac{\pi}{2})^2$,
where $r, \epsilon > 0$.

Embedded hyperplane hashing
transforms the database vector
(the normal of the query hyperplane)
into a high-dimensional vector,
\begin{align}
\bar{\mathbf{a}}
= \operatorname{vec}(\mathbf{a} \mathbf{a}^T)
[a_1^2, a_1a_2, \cdots, a_1a_d, a_2a_1, a_2^2,a_2a_3, \cdots, a_d^2].
\end{align}
Assuming $\mathbf{a}$
and $\mathbf{b}$ to be unit vectors,
the Euclidean distance between the embeddings
$\bar{\mathbf{a}}$ and $-\bar{\mathbf{b}}$
is given
$\|\bar{\mathbf{a}} - (-\bar{\mathbf{a}})\|_2^2
= 2 + 2 (\mathbf{a}^T\mathbf{b})^2$,
which means that
minimizing the distance
between the two embeddings
is equivalent to
minimizing $|\mathbf{a}^T\mathbf{b}|$.

The embedded hyperplane hash function family
is defined as
\begin{equation}
h(\mathbf{z}) = \left\{ \begin{array}{l l}
     h_{\mathbf{u}}(\bar{\mathbf{z}})  & \text{if $\mathbf{z}$ is a database vector}\\
     h_{\mathbf{u}}(-\bar{\mathbf{z}})  & \text{if $\mathbf{z}$ is a query hyperplane normal.}
   \end{array} \right.
\end{equation}
It is shown to be
$(r, r(1 + \epsilon), \frac{1}{\pi}\cos^{-1}\sin^2(\sqrt{r}), \frac{1}{\pi}\cos^{-1}\sin^2(\sqrt{r(1+\epsilon)}) )$
for
the angle distance $d_{\theta}(\mathbf{x}, \mathbf{n}) = (\theta_{\mathbf{x}, \mathbf{n}} - \frac{\pi}{2})^2$,
where $r, \epsilon > 0$.

It is also shown that
the exponent for embedded hyperplane hashing
is similar to that for XOR $1$-bit hyperplane hashing
and stronger than that for hyperplane hashing.

\subsection{Hamming Distance}
One LSH function for the Hamming distance
with binary vectors
$\mathbf{y} \in \{0, 1\}^d$
is proposed in~\cite{IndykM98},
$h(\mathbf{y}) = y_k$,
where $k \in \{1, 2, \cdots, d\}$
is a randomly-sampled index.
It can be shown that
$P(h(\mathbf{y}_i) = h(\mathbf{y}_j))
= 1 - \frac{\|\mathbf{y}_i - \mathbf{y}_j\|_h}{d}$.
It is proven that
the exponent $\rho$ is $1/c$.

\subsection{Jaccard Coefficient}
\subsubsection{Min-hash}
The Jaccard coefficient,
a similarity measure
between two sets,
$\mathcal{A}, \mathcal{B} \in \mathcal{U}$,
is defined
as $\operatorname{sim}(\mathcal{A}, \mathcal{B})
=\frac{\|\mathcal{A}\cap \mathcal{B}\|}{\|\mathcal{A} \cup \mathcal{B}\|}$.
Its corresponding distance is taken
as $1 - \operatorname{sim}(\mathcal{A}, \mathcal{B})$.
Min-hash~\cite{Broder97, BroderGMZ97}
is an LSH function for the Jaccard similarity.
Min-hash is defined as follows:
pick a random permutation $\pi$ from the ground universe $\mathcal{U}$,
and define $h(\mathcal{A}) = \min_{a \in \mathcal{A}} \pi(a)$.
It is easily shown that
$P(h(\mathcal{A}) = h(\mathcal{B})) = \operatorname{sim}(\mathcal{A}, \mathcal{B})$.
Given the $K$ hash values of two sets, the Jaccard similarity is estimated
as $\frac{1}{K}\sum_{k=1}^K \delta[h_k(\mathcal{A}) = h_k(\mathcal{B})]$,
where each $h_k$ corresponds to a random permutation that is independently generated.

\subsubsection{$K$-min sketch}
$K$-min sketch~\cite{Broder97, BroderGMZ97} is a generalization of min-wise sketch
(forming the hash values
using the $K$ smallest nonzeros from one permutation)
used for min-hash. It
also provides an unbiased estimator of
the Jaccard coefficient
but with a smaller variance,
which however cannot be used for approximate nearest neighbor search using hash tables
like min-hash.
Conditional random sampling~\cite{LiCH06, LiC07}
also takes the $k$ smallest nonzeros from one permutation,
and is shown to be a more accurate similarity estimator.
One-permutation hashing~\cite{LiOZ12},
also uses one permutation,
but breaks the space into $K$ bins,
and stores the smallest nonzero position in each bin
and concatenates them together to generate a sketch.
However, it is not directly applicable
to nearest neighbor search
by building hash tables
due to empty bins.
This issue is solved
by performing rotation
over one permutation hashing~\cite{ShrivastavaL14}.
Specifically,
if one bin is empty,
the hashed value from the first non-empty bin on the right (circular)
is borrowed as the key of this bin,
which supplies
an unbiased estimate of the resemblance
unlike~\cite{LiOZ12}.

\subsubsection{Min-max hash}
Min-max hash~\cite{JiLYTZ13},
instead of keeping the smallest hash value of each random permutation,
keeps both the smallest and largest values of each random permutation.
Min-max hash can generate $K$ hash values, using $\frac{K}{2}$ random permutations,
while still providing an unbiased estimator of the Jaccard coefficient, with a slightly smaller variance than min-hash.

\subsubsection{B-bit minwise hashing}
B-bit minwise hashing~\cite{LiKG10a, LiK10b}
only uses the lowest b-bits of the min-hash value
as a short hash value,
which gains substantial advantages in terms of storage space
while still leading to
an unbiased estimator of the resemblance
(the Jaccard coefficient).

\subsubsection{Sim-min-hash}
Sim-min-hash~\cite{ZhaoJG13} extends
min-hash to compare sets of real-valued vectors.
This approach first quantizes the real-valued vectors
and assigns an index (word)
for each real-valued vector.
Then, like the conventional min-hash, several random permutations
are used to generate the hash keys.
The different thing is that
the similarity is estimated
as
$\frac{1}{K}\sum_{k=1}^K \operatorname{sim}(\mathbf{x}^A_k, \mathbf{x}^B_k)$,
where $\mathbf{x}^A_k$ ($\mathbf{x}^B_k$ ) is the real-valued vector
(or Hamming embedding)
that is assigned to the word $h_k(\mathcal{A})$
($h_k(\mathcal{B})$),
and $\operatorname{sim}(\cdot, \cdot)$
is the similarity measure.

\subsection{$\chi^2$ Distance}
$\chi^2$-LSH~\cite{GorisseCP12} is a locality sensitive hashing function
for the $\chi^2$ distance.
The $\chi^2$ distance
over two vectors $\mathbf{x}_i$ and $\mathbf{x}_j$
is defined
as
\begin{align}
\chi^2(\mathbf{x}_i, \mathbf{x}_j) = \sqrt{\sum_{t=1}^d \frac{(x_{it} - x_{jt})^2}{x_{it} - x_{jt}}}.
\end{align}
The $\chi^2$ distance can also be defined without the square-root,
and the below developments still hold by substituting $r$
to $r^2$ in all the equations.

The $\chi^2$-LSH function
is defined as
\begin{align}
h_{\mathbf{w}, b}(\mathbf{x}) = \lfloor g_r(\mathbf{w}^T\mathbf{x}) + b \rfloor,
\end{align}
where $g_r(x) = \frac{1}{2}(\sqrt{\frac{8x}{r^2} + 1} - 1)$,
each entry of $\mathbf{w}$ is drawn from a $2$-stable distribution,
and $b$ is drawn from a uniform distribution over $[0, 1]$.

It can be shown that
\begin{align}
&~P(h_{\mathbf{w}, b}(\mathbf{x}_i) = h_{\mathbf{w}, b}(\mathbf{x}_j)) \nonumber \\
=&~\int_0^{(n+1)r^2} \frac{1}{c}f(\frac{t}{c})(1 - \frac{t}{(n+1)r^2}) dt,
\end{align}
where $f(t)$ denotes the probability density function
of the absolute value of the $2$-stable distribution,
$c = \|\mathbf{x}_i - \mathbf{x}_j\|_2$.

Let $c' = \chi^2(\mathbf{x}_i, \mathbf{x}_j)$.
It can be shown that
$P(h_{\mathbf{w}, b}(\mathbf{x}_i) = h_{\mathbf{w}, b}(\mathbf{x}_j))$
decreases monotonically with respect to $c$ and $c'$.
Thus, we can show it belongs to the LSH family.


\subsection{Other Similarities}
\subsubsection{Rank similarity}
Winner Take All (WTA) hash~\cite{YagnikSRL11}
is a sparse embedding method
that transforms the input feature space
into binary codes
such that the Hamming distance in the resulting space
closely correlates with rank similarity measure.
The rank similarity measure is shown to be more useful
for high-dimensional features
than the Euclidean distance,
in particular in the case of normalized feature vectors (e.g., the $\ell_2$ norm is equal to $1$).
The used similarity measure is a pairwise-order function,
defined as
\begin{align}
\operatorname{sim}_{po}(\mathbf{x}_1, \mathbf{x}_2)~&=~
\sum_{i=0}^{d-1}\sum_{j=1}^i \delta[ (x_{1i} - x_{1j})(x_{2i} - x_{2j})>0]\\
&=~\sum_{i=1}^d R_i(\mathbf{x}_1, \mathbf{x}_2),
\end{align}
where $R_i(\mathbf{x}_1, \mathbf{x}_2) = |L(\mathbf{x}_1, i) \cap L(\mathbf{x}_2, i)|$
and $L(\mathbf{x}_1, i) = \{j | x_{1i} > x_{1j}\}$.

WTA hash
generates a set of $K$ random permutations $\{\pi_k\}$.
Each permutation $\pi_k$
is used to reorder the elements of $\mathbf{x}$, yielding a new vector $\bar{\mathbf{x}}$.
The $k$th hash code
is computed as $\arg\max_{i=1}^T\bar{x}_i$,
taking a value between $0$ and $T-1$.
The final hash code is a concatenation of
$T$ values each corresponding to a permutation.
It is shown that WTA hash codes satisfy the LSH property
and min-hash is a special case of WTA hash.

\subsubsection{Shift invariant kernels}
Locality sensitive binary coding using shift invariant kernel hashing~\cite{RaginskyL09}
exploits the property
that the binary mapping of the original data
is guaranteed
to preserve the value of a shift-invariant kernel,
the random Fourier features (RFF)~\cite{RahimiR07}.
The RFF is defined as
\begin{align}
\phi_{\mathbf{w}, b}(\mathbf{x})
= \sqrt{2} \cos(\mathbf{w}^T\mathbf{x} + b),
\end{align}
where $\mathbf{w} \sim P_{K}$
and $b \sim \operatorname{Unif}[0, 2\pi]$.
For example,
for the Gaussian Kernel
$K(s) = e^{-\gamma \|s\|^2/2}$,
$\mathbf{w} \sim \operatorname{Normal}(\mathbf{0}, \gamma \mathbf{I})$.
It can be shown that
$\operatorname{E}_{\mathbf{w}, b}[\phi_{\mathbf{w}, b}(\mathbf{x}) \phi_{\mathbf{w}, b}(\mathbf{y})] = K(\mathbf{x}, \mathbf{y})$.

The binary code is computed
as
\begin{align}
\operatorname{sign}(\phi_{\mathbf{w}, b}(\mathbf{x}) + t),
\end{align}
where $t$ is a random threshold,
$t \sim \operatorname{Unif}[-1, 1]$.
It is shown that
the normalized Hamming distance
(i.e., the Hamming distance
divided by the number of bits
in the code string)
are both lower bounded and upper bounded
and that the codes preserve the similarity
in a probabilistic way.

\subsubsection{Non-metric distance}
Non-metric LSH~\cite{MuY10}
extends LSH to non-metric data
by embedding the data in the original space
into an implicit reproducing kernel Kre\u{i}n space
where the hash function is defined.
The kre\u{i}n space
with the indefinite inner product $<\cdot, \cdot>_{\mathcal{K}}$
$\mathcal{K}$
admits an orthogonal decomposition
as a direct sum
$\mathcal{K} = \mathcal{K}_+ \oplus \mathcal{K}_-$,
where
$(\mathcal{K}_+, \kappa_+(\cdot, \cdot))$
and $(\mathcal{K}_-, \kappa_-(\cdot, \cdot))$
are separable Hilbert spaces
with their corresponding positive definite inner products.
The inner product $\mathcal{K}$
is then computed as
\begin{align}
<\xi_+  +  \xi_-, \xi'_+ + \xi'_- >_{\mathcal{K}}
= \kappa_+(\xi_+, \xi'_+) - \kappa_-(\xi_-, \xi'_-).
\end{align}

Given the orthogonality
of $\mathcal{K}_+$ and $\mathcal{K}_-$of,
the pairwise $\ell_2$ distance in $\mathcal{K}$
is compute as
\begin{align}
\|\xi - \xi'\|_{\mathcal{K}}^2
= \|\xi_+ - \xi'_+\|_{\mathcal{K}_+}^2
- \|\xi_- - \xi'_-\|_{\mathcal{K}_-}^2.
\end{align}

The projections with the definite inner product
${\mathcal{K}_+}$ and ${\mathcal{K}_-}$
can be computed
using the technology in kernel LSH,
denoted by $p_+$
and $p_-$, respectively.
The hash function
with the input being
$(p_+(\xi) - p_+(\xi), p_+(\xi) + p_+(\xi)) = (a_1(\xi), a_2(\xi))$
and the output
being two binary bits
is defined as,
\begin{align}
\mathbf{h}(\xi) = [\delta[a_1(\xi)> \theta], \delta[a_2(\xi) > \theta]],
\end{align}
where $a_1(\xi)$ and $a_2(\xi)$ are assumed
to be normalized to $[0,1]$
and $\theta$ is a real number
uniformly drawn from $[0, 1]$.
It can be shown
that $P(\mathbf{h}(\xi) = \mathbf{h}(\xi'))
= (1 - |a_1(\xi) - a_1(\xi')|)(1-  |a_2(\xi) - a_2(\xi')|)$,
which indicates that the hash function belongs to the LSH family.

\subsubsection{Arbitrary distance measures}
The basic idea of distance-based hashing~\cite{AthitsosPPK08}
uses a line projection function
\begin{align}
&~f(\mathbf{x}; \mathbf{a}_1, \mathbf{a}_2) \nonumber \\
=&~\frac{1}{2 \operatorname{dist}(\mathbf{a}_1, \mathbf{a}_2)} (\operatorname{dist}^2(\mathbf{x}, \mathbf{a}_1) + \operatorname{dist}^2(\mathbf{a}_1, \mathbf{a}_2) - \operatorname{dist}^2(\mathbf{x}, \mathbf{a}_2)),
\end{align}
to formulate a hash function,
\begin{equation}
h(\mathbf{x}; \mathbf{a}_1, \mathbf{a}_2) = \left\{ \begin{array}{l l}
     1 & \quad \text{if $f(\mathbf{x}; \mathbf{a}_1, \mathbf{a}_2) \in [t_1, t_2]$}\\
     0 & \quad \text{otherwise.}
   \end{array} \right.
\end{equation}
Here, $\mathbf{a}_1$ and $\mathbf{a}_2$
are randomly selected data items,
$\operatorname{dist}(\cdot, \cdot)$
is the distance measure,
and $t_1$ and $t_2$ are two thresholds,
selected so that half of the data items are hashed to $1$
and the other half to $0$.

Similar to LSH,
distance-based hashing generates a compound hash function
using $K$ distance-based hash functions
and accordingly $L$ compound hash functions,
yielding $L$ hash tables.
However, it cannot be shown that
the theoretic guarantee in LSH holds for DBH.
There are some other schemes discussed in~\cite{AthitsosPPK08},
including optimizing $L$ and $K$ from the dataset,
applying DBH hierarchically so that
different set of queries use different parameters $L$ and $K$,
and so on.

\section{Locality Sensitive Hashing: Search, Modeling, and Analysis}
\label{sec:LSH2}

\subsection{Search}
\subsubsection{Entropy-based search}
The entropy-based search algorithm~\cite{Panigrahy06},
given a query point $\mathbf{q}$,
picks a set of ($O(N^{\rho})$) random points
$\mathbf{v}$
from $B(q, R)$,
a ball centered at $\mathbf{q}$
with the radius $r$
and searches in the buckets $H(\mathbf{v})$,
to find $cR$-near neighbors.
Here $N$ is the number of the database items,
$\rho = \frac{E}{\log (1/g)}$,
$M$
is the entropy
$I(h(\mathbf{p}) | \mathbf{q}, R)$
where $\mathbf{p}$ is a random point in $B(q, R)$,
and $g$ denotes the upper bound
on the probability
that two points that are
at least distance $cr$ apart
will be hashed to the same bucket.
In addition,
the search algorithm suggests
to build a single hash table
with $K = \frac{N}{\log (1/g)}$ hash bits.

The paper~\cite{Panigrahy06}
presents the theoretic evidence
theoretically guaranteeing the search quality.


\subsubsection{LSH forest}
LSH forest~\cite{BawaCG05}
represents each hash table,
built from LSH,
using a tree,
by pruning subtrees (nodes) that
do not contain any database points
and also restricting the depth of each leaf node
not larger than a threshold.
Different from the conventional scheme
that finds the candidates
from the hash buckets corresponding to the hash codes
of the query point,
the search algorithm
finds the points
contained in
subtrees over LSH forest having the largest prefix match
by a two-phase approach:
the first top-down phase descends each LSH tree
to find the leaf having the largest prefix match
with the hash code of the query,
the second bottom-up phase
back-tracks each tree from the discovered leaf nodes in the first phase
in the largest-prefix-match-first manner
to find
subtrees having the largest prefix match
with the hash code of the query.


\subsubsection{Adaptative LSH}
The basic idea of adaptative LSH~\cite{JegouASG08}
is to select the most relevant hash codes
based on the relevance value.
The relevance value
is computed
by accumulating the differences between the projection value
and the mean of the corresponding line segment along the projection direction
(or equivalently the difference of the projection values along the projection directions
and the center of the corresponding bucket).


\subsubsection{Multi-probe LSH}
The basic idea of multi-probe LSH~\cite{LvJWCL07}
is to intelligently probe
multiple buckets that are likely to contain query results in a hash table,
whose hash values may not necessarily be the same to the hash value of the query vector.
Given a query $\mathbf{q}$,
with its hash code denoted by $g(\mathbf{q}) = (h_1(\mathbf{q}), h_2(\mathbf{q}), \cdots, h_K(\mathbf{q}))$,
multi-probe LSH finds a sequence of hash perturbation vector,
$\{\boldsymbol{\delta}_i = \{\delta_{i1}, \delta_{i2}, \cdots, \delta_{iK}\}\}$
and sequentially probe the hash buckets
$\{g(\mathbf{q})  + \boldsymbol{\delta}_{o(i)}\}$.
A score, computed
as $\sum_{j=1}^Kx_j^2(\delta_{ij})$,
where $x_j(\delta_{ij})$
is the distance of $\mathbf{q}$
from the boundary of the slot $h_j(\mathbf{q}) + \delta_j$,
is used to sort the perturbation vectors,
so that the buckets are accessed
in order of increasing the scores.
The paper~\cite{LvJWCL07}
also proposes to use
the expectation $\operatorname{E}(x_j^2(\delta_{ij}))$,
which is estimated with the assumption
that $\delta_{ij}$ is uniformly distributed in $[0, r]$
($r$ is the width of the hash function used for Euclidean LSH),
to replace $x_j^2 (\delta_{ij})$
for sorting the perturbation vectors.
Compared with conventional LSH,
to achieve the same search quality,
multi-probe LSH has a similar time efficiency
while reducing the number of hash tables by an order of magnitude.

The posteriori multi-probe LSH algorithm presented in~\cite{JolyB08}
gives a probabilistic interpretation
of multi-probe LSH
and presents a probabilistic score,
to sort the perturbation vectors.
The basic ideas of the probabilistic score computation
include
the property (likelihood) that
the difference of the projections of two vectors
along a random projection direction
drawn from a Gaussian distribution
follows a Gaussian distribution,
as well as
estimating the distribution (prior)
of the neighboring points of a point
from the train query points and their neighboring points
with assuming that
the neighbor points of a query point
follow a Gaussian distribution.


\subsubsection{Dynamic collision counting for search}
The collision counting LSH scheme introduced in~\cite{GanFFN12}
uses a base of $m$ single hash functions
to construct dynamic compound hash functions,
instead of $L$ static compound hash functions
each of which is composed of $K$ hash functions.
This scheme regards a data vector
that collides with the query vector
over at least $K$ hash functions out of the base of $m$ single hash functions
as a good cR-NN candidate.
The theoretical analysis shows
that such a scheme
by appropriately choosing $m$ and $K$
can have a guarantee on search quality.
In case that there is no data returned for a query
(i.e., no data vector has at least $K$ collisions with the query),
a virtual reranking scheme is presented
with the essential idea of
expanding the window width
gradually
in the hash function for E2LSH,
to increase the collision chance,
until finding enough number of data vectors
that have at least $K$ collisions with the query.

\subsubsection{Bayesian LSH}
The goal of Bayesian LSH~\cite{SatuluriP12}
is to estimate the probability distribution,
$p(s|M(m,k))$,
of the true similarity $s$
in the case that $m$ matches out of $k$ hash bits
for a pair of hash codes $(g(\mathbf{q}), g(\mathbf{p}))$
of the query vector $\mathbf{q}$ and a NN candidate $\mathbf{p}$,
which is denoted by $M(m,k)$,
and prune the candidate $\mathbf{p}$
if the probability for the case $s \geqslant t$
with $t$ being a threshold
is less than $\epsilon$.
In addition, if the concentration probability $P(|s - s^*| \leqslant \delta
| M(m,k)) \geqslant \lambda$,
or intuitively the true similarity $s$
under the distribution $p(s|M(m,k))$
is almost located near the mode,
$s^* = \arg\max_s p(s|M(m,k))$,
the similarity evaluation is early stopped
and such a pair is regarded
as similar enough,
which is an alternative
of computing the exact similarity of such a pair
in the original space.
The paper~\cite{SatuluriP12}
gives two examples
of Bayesian LSH for the Jaccard similarity
and the arccos similarity
for which $p(s|M(m,k))$ are instantiated.

\subsubsection{Fast LSH}
Fast LSH~\cite{DasguptaKS11}
presents two algorithms,
ACHash and DHHash,
that formulate $L$ $K$-bits compound hash functions.
ACHash pre-conditions the input vector
using a random diagonal matrix
and a Hadamard transform,
and then applies a sparse Gaussian matrix followed by a rounding.
DHHash does the same pre-conditioning process
and then applies a random permutation,
followed by a random diagonal Gaussian matrix
and an another Hadamard transform.
It is shown that
it takes
only $O(d\log d + KL)$
for both ACHash and DHHash
to compute hash codes
instead of $O(dKL)$.
The algorithms are also extended
to the angle-based similarity,
where the query time to $\epsilon$-approximate
the angle between two vectors
is reduced
from $O(d/{\epsilon}^2)$
to $O(d \log 1/\epsilon + 1/{\epsilon}^2)$.

\subsubsection{Bi-level LSH}
The first level of bi-level LSH~\cite{PanM12}
uses a random-projection tree
to divide the dataset
into subgroups
with bounded aspect ratios.
The second level is an LSH table,
which is basically implemented
by randomly projecting data points into a low-dimensional space
and then partitioning the low-dimensional space
into cells.
The table is enhanced using a hierarchical structure.
The hierarchy, implemented
using the space filling Morton curve
(a.k.a., the Lebesgue or Z-order curve),
is useful when there are not enough candidates
retrieved for the multi-probe LSH algorithm.
In addition, the $E_8$ lattice is used
for partitioning the low-dimensional space
to overcome the curse of dimensionality
caused by the basic $Z^M$ lattice.

\subsection{SortingKeys-LSH}

SortingKeys LSH~\cite{LiuCHLS14}
aims at improving the search scheme of LSH by reducing random I/O operations when retrieving candidate data points.
The paper defines a distance measure between compound hash keys to estimate the true distance between data points,
and introduces a linear order on the set of compound hash keys.
The method sorts all the compound hash keys in ascending order and stores the corresponding data points on disk according to this order,
then close data points are likely to be stored locally.
During ANN search, a limited number of pages on the disk, which are ``close'' to the query in terms of the distance defined between compound hash keys, are needed to be accessed for sufficient candidate generation,
leading to much shorter response time due to the reduction of random I/O operations, yet with higher search accuracy.

\subsection{Analysis and Modeling}
\subsubsection{Modeling LSH}
The purpose~\cite{DongWJCL08}
is to model the recall and the selectivity
and apply it to determine the optimal parameters,
the window size $r$,
the number of hash functions $K$
forming the compound hash function,
the number of tables $L$,
and the number of bins $T$ probed in each table
for $E2$LSH.
The recall is defined as
the percentage of the true NNs
in the retrieved NN candidates.
The selectivity is defined as
the ratio of the number of the retrieved candidates
to the number of the database points.
The two factors
are formulated as
a function of the data distribution,
for which the squared $L_2$ distance
is assumed to follow a Gamma distribution
that is estimated from the real data.
The estimated distributions of $1$-NN, $2$-NNs, and so on
are used to compute the recall and selectivity.
Finally, the optimal parameters
are computed
to minimize the selectivity
with the constraint
that the recall is not less than a required value.
A similar and more complete analysis for parameter optimization
is given in~\cite{SlaneyLH12}

\subsubsection{The difficulty of nearest neighbor search}
\cite{HeKC12} introduces a new measure,
relative contrast for analyzing the meaningfulness
and difficulty of nearest neighbor search.
The relative contrast for a query $\mathbf{q}$,
given a dataset $\mathbf{X}$
is defined as
$C^q_r = \frac{\operatorname{E}_{\mathbf{x}}[d(\mathbf{q}, \mathbf{x})]}{\min_{\mathbf{x}}(d(\mathbf{q}, \mathbf{x}))}$.
The relative contrast expectation with respect to the queries
is given as follows,
$C_r = \frac{\operatorname{E}_{\mathbf{x}, \mathbf{q}}[d(\mathbf{q}, \mathbf{x})]}{\operatorname{E}_{\mathbf{q}}[\min_{\mathbf{x}}(d(\mathbf{q}, \mathbf{x}))]}$.

Define a random variable $R = \sum_{j=1}^d R_j = \sum_{j=1}^d\operatorname{E}_{\mathbf{q}}[\|x_j - q_j\|_p^p]$,
and let the mean be $\mu$ and the variance be $\sigma^2$.
Define the normalized variance: $\sigma'^2 = \frac{\sigma^2}{\mu^2}$.
It is shown that
if $\{R_1, R_2, \cdots, R_d\}$ are independent
and satisfy Lindeberg's condition,
the expected relative contrast is approximated as,
\begin{align}
C_r \approx \frac{1}{[1 + \phi^{-1} (\frac{1}{N} + \phi(\frac{-1}{\sigma'}))\sigma']^{\frac{1}{p}}},
\end{align}
where $N$ is the number of database points,
$\phi(\cdot)$ is the cumulative density function of standard Gaussian,
$\sigma'$ is the normalized standard deviation,
and $p$ is the distance metric norm.
It can also be generalized to the relative contrast for the $k$th nearest neighbor,
\begin{align}
C_r^k = \frac{\operatorname{E}_{\mathbf{x}, \mathbf{q}}[d(\mathbf{q}, \mathbf{x})]}{\operatorname{E}_{\mathbf{q}}[\operatorname{k-min}_{\mathbf{x}}(d(\mathbf{q}, \mathbf{x}))]}
\approx \frac{1}{[1 + \phi^{-1} (\frac{k}{N} + \phi(\frac{-1}{\sigma'}))\sigma']^{\frac{1}{p}}},
\end{align}
where $\operatorname{k-min}_{\mathbf{x}}(d(\mathbf{q}, \mathbf{x}))$ is the distance of the query
to the $k$th nearest neighbor.

Given the approximate relative contrast,
it is clear
how the data dimensionality $d$, the database size $N$,
the metric norm $p$,
and the sparsity of the data vector (determining $\sigma'$) influence the relative contrast.

It is shown that
LSH, under the $\ell_p$-norm distance, can find the exact nearest neighbor with probability $1-\delta$
by returning $O(\log \frac{1}{\delta} n^{g(C_r)})$
candidate points,
where $g(C_r)$ is a function monotonically decreasing with $C_r$,
and that,
in the context of linear hashing $\operatorname{sign}(\mathbf{w}^T\mathbf{x} + b)$,
the optimal projection that maximizes the relative contrast
is $\mathbf{w}* = \arg\max_{\mathbf{w}} \frac{\mathbf{w}^T\boldsymbol{\Sigma}_x\mathbf{w}}{\mathbf{w}^T\mathbf{S}_{NN}\mathbf{w}}$,
where $\boldsymbol{\Sigma}_x = \frac{1}{N}\sum_{i=1}^N \mathbf{x}_i \mathbf{x}_i^T$
and $\mathbf{S}_{NN} = \operatorname{E}_{\mathbf{q}} [(\mathbf{q} - \operatorname{NN}(\mathbf{q}))(\mathbf{q} - \operatorname{NN}(\mathbf{q}))^T]$,
subject to $\mathbf{S}_{NN} = \mathbf{I}$,
$\mathbf{w}* = \arg\max_{\mathbf{w}} \mathbf{w}^T\boldsymbol{\Sigma}_x\mathbf{w}$.

The LSH scheme has very nice theoretic properties.
However,
as the hash functions are data-independent,
the practical performance is not
as good as expected in certain applications.
Therefore,
there are a lot of followups
that learn hash functions from the data.

\section{Learning to Hash: Hamming Embedding and Extensions}
\label{sec:LTH1}
Learning to hash
is a task of
learning a compound hash function,
$\mathbf{y} = \mathbf{h}(\mathbf{x})$,
mapping an input item $\mathbf{x}$
to a compact code $\mathbf{y}$,
such that the nearest neighbor search
in the coding space is efficient
and the result is an effective approximation
of the true nearest search result in the input space.
An instance
of the learning-to-hash approach
includes three elements:
hash function,
similarity measure in the coding space,
and optimization criterion.
Here
The similarity in similarity measure
is a general concept,
and may mean distance or other forms of similarity.

\vspace{.1cm}
\noindent\textbf{Hash function.}
The hash function can be based on linear projection,
spherical function,
kernels,
and neural network,
even a non-parametric function,
and so on.
One popular hash function is a linear hash function:
$y = \operatorname{sign}(\mathbf{w}^T\mathbf{x}) \in \{0, 1\}$,
where $\operatorname{sign}(\mathbf{w}^T\mathbf{x}) = 1$
if $\mathbf{w}^T\mathbf{x} \geqslant 0$
and $\operatorname{sign}(\mathbf{w}^T\mathbf{x}) = 0$
otherwise.
Another widely-used hash function is
a function based on nearest vector assignment:
$y = \arg\min_{k \in \{1, \cdots, K\}} \|\mathbf{x} - \mathbf{c}_k\|_2 \in \mathbb{Z}$,
where $\{\mathbf{c}_1, \cdots, \mathbf{c}_K\}$ is a set of centers,
computed by some algorithm, e.g., $K$-means.

The choice of hash function types
influences the efficiency of computing hash codes
and the flexility of the hash codes,
or the flexibility of partitioning the space.
The optimization of hash function parameters
is dependent to
both distance measure and distance preserving.

\vspace{.1cm}
\noindent\textbf{Similarity measure.}
There are two main distance measure schemes in the coding space.
Hamming distance with its variants,
and Euclidean distance.
Hamming distance is widely used
when the hashing function maps the data point
into a Hamming code $\mathbf{y}$
for which each entry is either $1$
or $0$,
and is defined as
the number of bits
at which the corresponding values are different.
There are some other variants,
such as weighted Hamming distance, distance table lookup,
and so on.
Euclidean distance is used
in the approaches based on nearest vector assignment
and evaluated between the vectors
corresponding to the hash codes,
i.e., the nearest vectors assigned to the data vectors,
which is efficiently computed
by looking up a precomputed distance table.
There is a variant,
asymmetric Euclidean distance,
for which only one vector is approximated by its nearest vector
while the other vector is not approximated.
There are also some works
learning a distance table between hash codes
by assuming the hash codes are already given.

\vspace{.1cm}
\noindent\textbf{Optimization criterion.}
The approximate nearest neighbor search result
is evaluated by comparing it
with the true search result,
that is the result according to the distance
computed in the input space.
Most similarity preserving criteria
design various forms
as the surrogate
of such an evaluation.

The straightforward form is
to directly compare the order of
the ANN search result
with that of the true result
(using the reference data points as queries),
which called the order-preserving criterion.
The empirical results show
that the ANN search result usually has higher probability
to approach the true search result
if the distance computed in the coding space
accurately approximates the distance computed in the input space.
This motivates the so-called similarity alignment criterion,
which directly minimizes the differences
between the distances (similarities) computed in the coding and input space.
An alternative surrogate is
coding consistent hashing,
which penalizes the larger distances in the coding space
but with the larger similarities in the input space (called coding consistent to similarity, shorted as coding consistent
as a major of algorithms use it)
and encourages the smaller (larger) distances in the coding space
but with the smaller (larger) distances in the input space
(called coding consistent to distance).
One typical approach,
the space partitioning approach,
assumes
that space partitioning has already implicitly
preserved the similarity
to some degree.

Besides similarity preserving,
another widely-used criterion
is coding balance,
which means that the reference vectors
should be uniformly distributed in each bucket
(corresponding to a hash code).
Other related criteria,
such as bit balance,
bit independence,
search efficiency,
and so on,
are essentially (degraded) forms
of coding balance.


\begin{table}
\centering
\caption{Hash functions}
\label{tab:lth:hashfunctiontypes}
\begin{tabular}{|l|c|}
\hline
 type &  abbreviation \\
\hline
\hline
linear &  LI \\
 bilinear &  BILI \\
 Laplacian eigenfunction &   LE \\
kernel &   KE \\
quantizer &  QU \\
$1$D quantizer  & OQ \\
spline & SP \\
neural network  & NN \\
spherical function & SF \\
classifier &  CL \\
\hline
\end{tabular}
\end{table}


\begin{table}[t]
\centering
\caption{Distance measures in the coding space}
\label{tab:lth:distancemeasures}
\begin{tabular}{|c|c|}
\hline
type & abbreviation \\
\hline
\hline
Hamming distance & HD \\
normalized Hamming distance &  NHD \\
asymmetric Hamming distance & AHD \\
weighted Hamming distance & WHD \\
query-dependent weighted Hamming distance &  QWHD \\
normalized Hamming affinity & NHA \\
\hline
Manhattan & MD \\
\hline
asymmetric Euclidean distance  & AED \\
symmetric Euclidean distance  & SED \\
\hline
lower bound & LB\\
\hline
\end{tabular}
\end{table}

\begin{table}[t]
\centering
\caption{Optimization criterion.}
\label{tab:lth:optimizationcriterion}
\begin{tabular}{|l|c|}
\hline
type & abbreviation \\
\hline
\hline
\multicolumn{2}{|l|}{Hamming embedding} \\
\hline
coding consistent & CC \\
coding consistent to distance & CCD \\
code balance & CB \\
bit balance & BB \\
bit uncorrelation & BU \\
projection uncorrelation & PU \\
mutual information maximization & MIM \\
\hline
minimizing differences between distances & MDD \\
minimizing differences between similarities & MDS \\
minimizing differences between similarity distribution & MDSD\\
\hline
hinge-like loss & HL \\
rank order loss & ROL \\
triplet loss & TL \\
classification error & CE \\
\hline
space partitioning & SP \\
complementary partitioning & CP \\
pair-wise bit balance & PBB \\
maximum margin & MM \\
\hline
\multicolumn{2}{|l|}{Quantization} \\
\hline
bit allocation & BA \\
quantization error & QE \\
equal variance & EV \\
maximum cosine similarity & MCS \\
\hline
\end{tabular}
\end{table}

\begin{table*}[t]
\centering
\caption{A summary of hashing algorithms. $^*$ means that
hash function learning does not explicitly rely on the distance measure in the coding space.
S = semantic similarity.
E = Euclidean distance.
sim. = similarity.
dist. = distance.}
\footnotesize
\label{tab:lth:summaryofhashingalgorithms}
\begin{tabular}{|l||c||c|c|c|}
\hline
method & input sim. & hash function & dist. measure & optimization criteria\\
\hline
\hline
spectral hashing~\cite{WeissTF08} & E & LE & HD & CC + BB + BU \\
kernelized spectral hashing~\cite{HeLC10}& S, E & KE & HD & CC + BB + BU \\
Hypergraph spectral hashing~\cite{ZhuangLWZS11, LiuSXWZ13}& S & CL & HD & CC + BB + BU \\
Topology preserving hashing~\cite{ZhangZTGLT13} & E & LI & HD & CC + CCD + BB + BU \\
hashing with graphs~\cite{LiuWKC11} & S & KE & HD & CC + BB \\
ICA Hashing~\cite{HeCRB11} & E & LI, KE & HD & CC + BB + BU + MIM \\
Semi-supervised hashing~\cite{WangKC10a, WangKC10b, WangKC12} & S, E & LI  & HD & CC + BB + PU \\
LDA hash~\cite{StrechaBBF12} & S & LI & HD & CC + PU\\
\hline
binary reconstructive embedding~\cite{KulisD09} & E & LI, KE & HD & MDD \\
supervised hashing with kernels~\cite{LiuWJJC12} & E, S & LI, KE & HD & MDS \\
spec hashing~\cite{LinRY10} & S & CL & HD & MDSD \\
bilinear hyperplane hashing~\cite{LiuWMKC12} & ACS & BILI & HD & MDS \\
\hline
minimal loss hashing~\cite{NorouziF11} & E, S & LI & HD & HL \\
order preserving hashing~\cite{WangWYL13} & E & LI & HD & ROL \\
Triplet loss hashing~\cite{NorouziFS12} & E, S & Any & HD, AHD & TL \\
listwise supervision hashing~\cite{WangLSJ13} & E, S & LI & HD & TL \\
Similarity sensitive coding (SSC)~\cite{Shakhnarovich05} & S & CL & WHD & CE \\
parameter sensitive hashing~\cite{ShakhnarovichVD03} & S & CL & WHD & CE \\
column generation hashing~\cite{LiLSHD13} & S & CL & WHD & CE \\
\hline
complementary projection hashing~\cite{JinHLZLCL13}$^*$ & E & LI, KE & HD & SP + CP + PBB \\
label-regularized maximum margin hashing~\cite{MuSY10}$^*$ & E, S & KE & HD & SP + MM + BB \\
Random maximum margin hashing~\cite{JolyB11}$^*$ & E & LI, KE & HD & SP + MM + BB \\
spherical hashing~\cite{HeoLHCY12}$^*$ & E & SF & NHD & SP + PBB \\
density sensitive hashing~\cite{LinCL12}$^*$ & E & LI & HD & SP + BB \\
\hline
multi-dimensional spectral hashing~\cite{WeissFT12} & E & LE & WHD &  CC + BB + BU \\
Weighted hashing~\cite{WangZS13} & E & LI & WHD & CC + BB + BU \\
\hline
Query-adaptive bit weights~\cite{JiangWC11, JiangWXC13} & S & LI (all) & QWHD & CE \\
Query adaptive hashing~\cite{LiuYJHZ13} & S & LI & QWHD & CE \\
\hline
\end{tabular}
\end{table*}

In the following,
we review Hamming bedding based hashing algorithms.
Table~\ref{tab:lth:summaryofhashingalgorithms}
presents the summary of the algorithms reviewed
from Section~\ref{sec:csh} to Section~\ref{sec:whd},
with some concepts given
in Tables~\ref{tab:lth:hashfunctiontypes},~\ref{tab:lth:hashfunctiontypes}
and~\ref{tab:lth:optimizationcriterion}.

\subsection{Coding Consistent Hashing}
\label{sec:csh}
Coding consistent hashing refers to a category of
hashing functions
based on minimizing the similarity weighted distance,
$s_{ij}d(\mathbf{y}_i, \mathbf{y}_j)$
(and possibly maximizing $d_{ij}d(\mathbf{y}_i, \mathbf{y}_j)$),
to formulate the objective function.
Here, $s_{ij}$ is the similarity between $\mathbf{x}_i$ and $\mathbf{x}_j$
computed from the input space or given from the semantic meaning.

\subsubsection{Spectral hashing}
Spectral hashing~\cite{WeissTF08},
the pioneering coding consistency hashing algorithm,
aims to find an easy-evaluated hash function
so that
(1) similar items are mapped to similar hash codes
based on the Hamming distance (coding consistency)
and (2) a small number of hash bits are required.
The second requirement is a form similar to coding balance,
which is transformed to two requirements:
bit balance
and bit uncorrelation.
The balance means that
each bit has around $50\%$ chance of being $1$ or $0$ ($-1$).
The uncorrelation means that different bits are uncorrelated.

Let $\{\mathbf{y}_n\}{n=1}^N$ be the hash codes of the $N$ data items,
each $\mathbf{y}_n$ be a binary vector of length $M$.
Let $s_{ij}$ be the similarity that correlates with the Euclidean distance.
The formulation is given as follows:

\begin{align}
\min_{\mathbf{Y}}~&~ \operatorname{Trace}(\mathbf{Y}(\mathbf{D}-\mathbf{S})\mathbf{Y}^T)   \\
\operatorname{s.t.}
~&~ \mathbf{Y}\mathbf{1} = \mathbf{0}   \\
~&~ \mathbf{Y} \mathbf{Y}^T = \mathbf{I} \\
~&~ y_{im} \in \{-1, 1\},
\label{eqn:spectralhashingobjectivefunction}
\end{align}
where $\mathbf{Y} = [\mathbf{y}_1 \mathbf{y}_2 \cdots \mathbf{y}_N]$,
$\mathbf{S}$ is a matrix $[s_{ij}]$ of size $N \times N$,
$\mathbf{D}$ is a diagonal matrix $\operatorname{Diag}(d_{11}, \cdots, d_{NN})$,
and $d_{nn} = \sum_{i=1}^N s_{ni}$.
$\mathbf{D}-\mathbf{S}$ is called Laplacian matrix
and $\operatorname{Trace}(\mathbf{Y}(\mathbf{D}-\mathbf{S})\mathbf{Y}^T)= \sum_{i=1}^N\sum_{j=1}^N w_{ij}\|\mathbf{y}_i - \mathbf{y}_j\|_2^2$.
$\mathbf{Y}\mathbf{1} = \mathbf{0}$ corresponds to the bit balance requirement.
$\mathbf{Y} \mathbf{Y}^T = \mathbf{I}$ corresponds to
the bit uncorrelation requirement.

Rather than solving the problem~Equation~\ref{eqn:spectralhashingobjectivefunction} directly,
a simple approximate solution with the assumption of uniform data distribution
is presented in~\cite{WeissTF08}.
The algorithm is given as follows:
\begin{enumerate}
  \item Find the principal components of the $N$ $d$-dimensional reference data items using principal component analysis (PCA).
  \item Compute the $M$ $1$D Laplacian eigenfunctions with the smallest eigenvalues
  along each PCA direction.
  \item Pick the $M$ eigenfunctions with the smallest eigenvalues among $Md$ eigenfunctions.
  \item Threshold the eigenfunction at zero, obtaining the binary codes.
\end{enumerate}

The $1$D Laplacian eigenfunction for
the case of uniform distribution on $[r_l, r_r]$
is $\phi_f(x) = \sin(\frac{\pi}{2} + \frac{f\pi}{r_r - r_l}x)$
and the corresponding eigenvalue is
$\lambda_f = 1 - \exp{(-\frac{\epsilon^2}{2} |\frac{f\pi}{r_r - r_l}|^2)}$,
where $f = 1, 2, \cdots$ is the frequency
and $\epsilon$ is a fixed small value.

The assumption
that
the data is uniformly distributed
does not hold in real cases,
resulting in that the performance
of spectral hashing is deteriorated.
Second,
the eigenvalue monotonously increases
with respect to $|\frac{f}{r_r - r_l}|^2$,
which means that
the PCA direction with a large spread ($|r_r - r_l|$)
and a lower frequency ($f$) is preferred.
This means that there might be more than one eigenfunctions picked along a single PCA direction,
which breaks the uncorrelation requirement,
and thus the performance is influenced.
Last,
thresholding the eigenfunction
$\phi_f(x) = \sin(\frac{\pi}{2} + \frac{f\pi}{r_r - r_l}x)$
at zero
leads to
that near points are mapped to different values
and even far points are mapped to the same value.
It turns out that
the hamming distance is not well consistent to the Euclidean distance.

In the case that the spreads along the top $M$ PCA direction are the same,
the spectral hashing algorithm
actually partitions each direction into two parts
using the median (due to the bit balance requirement) as the threshold.
It is noted that,
in the case of uniform distributions,
the solution is equivalent to thresholding at the mean value.
In the case that the true data distribution is a multi-dimensional isotropic Gaussian distribution,
it is equivalent to iterative quantization~\cite{GongL11, GongLGP13} and isotropic hashing~\cite{KongL12a}.

Principal component hashing~\cite{MatsushitaW09}
also uses the principal direction
to formulate the hash function.
Specifically,
it partitions the data points into $K$ buckets
so that the projected points along the principal direction
are uniformly distributed in the $K$ buckets.
In addition, bucket overlapping
is adopted to deal with the boundary issue
(neighboring points around the partitioning position are assigned
to different buckets).
Different from spectral hashing,
principal component hashing aims at
constructing hash tables
rather than compact codes.

The approach in~\cite{LiWCXL13}, spectral hashing with semantically consistent graph
first learns a linear transform matrix
such that
the similarities computed over the transformed space
is consistent to
the semantic similarity as well as the Euclidean distance-based similarity,
then applies spectral hashing to learn hash codes.

\subsubsection{Kernelized spectral hashing}
The approach introduced in~\cite{HeLC10} extends spectral hashing
by explicitly defining the hash function using kernels.
The $m$th hash function is given as follows,
\begin{align}
&~y_{m} = h_m(\mathbf{x}) \\
= &~\operatorname{sign}(\sum_{t=1}^{T_m} w_{mt} K(\mathbf{s}_{mt}, \mathbf{x}) - b_m)\\
= &~\operatorname{sign}(\sum_{t=1}^{T_m} w_{mt}<\phi(\mathbf{s}_{mt}), \phi(\mathbf{x})> - b_m)\\
= &~\operatorname{sign}(<\mathbf{v}_m, \phi(\mathbf{x})> - b_m).
\label{eqn:kerneklizedspectralhashfunction}
\end{align}
Here $\{\mathbf{s}_{mt}\}_{t=1}^{T_m}$ is the set of randomly-sampled anchor items
for forming the hash function,
and its size $T_m$ is usually the same for all $M$ hash functions.
$K(\cdot, \cdot)$ is a kernel function,
and $\phi(\cdot)$ is its corresponding mapping function.
$\mathbf{v}_m = [w_{m1}\phi(\mathbf{s}_{m1})~\cdots w_{mT_m}~\phi(\mathbf{s}_{mT_m})]^T$.

The objective function is written as:
\begin{align}
\min_{\{w_{mt}\}}~&~ \operatorname{Trace}(\mathbf{Y}(\mathbf{D}-\mathbf{S})\mathbf{Y}^T)  + \sum_{m=1}^M \|\mathbf{v}_m\|_2^2,
\end{align}
The constraints are the same to those of spectral hashing,
and differently the hash function is given in Equation~\ref{eqn:kerneklizedspectralhashfunction}.
To efficiently solve the problem,
the sparse similarity matrix $\mathbf{W}$ and
the Nystr\"{o}m algorithm
are used to reduce the computation cost.

\subsubsection{Hypergraph spectral hashing}
Hypergraph spectral hashing~\cite{ZhuangLWZS11, LiuSXWZ13}
extends spectral hashing
from an ordinary (pair-wise) graph
to a hypergraph (multi-wise graph),
formulates the problem
using the hypergraph Laplacian
(replace the graph Laplacian~\cite{WeissTF08, WeissFT12})
to form the objective function,
with the same constraints to spectral hashing.
The algorithm in~\cite{ZhuangLWZS11, LiuSXWZ13}
solves the optimization problem,
by relaxing the binary constraint
eigen-decomposing the the hypergraph Laplacian matrix,
and thresholding the eigenvectors at zero.
It computes the code for an out-of-sample vector,
by regarding each hash bit as a class label of the data vector
and learning a classifier for each bit.
In essence, this approach is a two-step approach
that separates the optimization of coding and hash functions.
The remaining challenge lies in
how to extend the algorithm to large scale
because the eigen-decomposition step is quite time-consuming.

\subsubsection{Sparse spectral hashing}

Sparse spectral hashing~\cite{ShaoWOZ12} combines sparse principal component analysis (Sparse PCA)
and Boosting Similarity Sensitive Hashing (Boosting SSC)
into traditional spectral hashing.
The problem is formulated as
as thresholding a subset of eigenvectors of the Laplacian graph
by constraining the number of nonzero features.
The convex relaxation makes the learnt codes globally optimal
and the out-of-sample extension is achieved
by learning the eigenfunctions.


\subsubsection{ICA hashing}
The idea of independent component analysis (ICA) Hashing~\cite{HeCRB11} starts from coding balance.
Intuitively coding balance means
that the average number of data items mapped
to each hash code is the same.
The coding balance requirement is formulated
as maximizing the entropy $\operatorname{entropy}(y_1, y_2, \cdots, y_M)$,
and subsequently formulated as
bit balance:
$\operatorname{E}(y_m) = 0$
and mutual information minimization:
$I(y_1, y_2, \cdots, y_M)$.

The approach approximates the mutual information
using the scheme similar to the one widely used independent component analysis.
The mutual information
is relaxed:
$I(y_1, y_2, \cdots, y_M) = I(\mathbf{w}_1^T\mathbf{x}, \mathbf{w}_2^T\mathbf{x}), \cdots, \mathbf{w}_M^T\mathbf{x})$
and is approximated as maximizing
\begin{align}
\sum_{m=1}^M \| c - \frac{1}{N}\sum_{n=1}^N g(\mathbf{W}^T\mathbf{x}_n) \|_2^2,
\end{align}
under the constraint of whiten condition (which can be derived from bit uncorrelation),
$\mathbf{w}_i^T\operatorname{E}(\mathbf{x}\mathbf{x}^T)\mathbf{w}_j = \delta[i
=j]$,
$c$ is a constant,
$g(u)$ is some non-quadratic functions,
such that $g(u) = - \exp{(-\frac{u^2}{2})}$
or $g(u) = \log\cosh(u)$.

The whole objective function together preserving the similarities as done in spectral hashing is
written as follows,
\begin{align}
\max_{\mathbf{W}}~&~ \sum_{m=1}^M \| c - \frac{1}{N}\sum_{n=1}^N g(\mathbf{W}^T\mathbf{x}_n) \|_2^2 \\
s.t.~&~ \mathbf{w}_i^T\operatorname{E}(\mathbf{x}\mathbf{x}^T)\mathbf{w}_j\ = \delta[i=j] \\
~&~ \operatorname{trace} (\mathbf{W}^T\boldsymbol{\Sigma}\mathbf{W}) \leq \eta.
\end{align}
The paper~\cite{HeCRB11} also presents a kernelized version
by using the kernel hash function.

%
%
%

\subsubsection{Semi-supervised hashing}
Semi-supervised hashing~\cite{WangKC10a, WangKC10b, WangKC12}
extends spectral hashing
into the semi-supervised case,
in which some pairs of data items
are labeled as belonging to the same semantic concept,
some pairs are labeled
as belonging to different semantic concepts.
Specifically,
the similarity weight $s_{ij}$ is assigned to
$1$ and $-1$
if the corresponding pair of data items,
$(\mathbf{x}_i, \mathbf{x}_j)$,
belong to the same concept,
and different concepts,
and $0$ if no labeling information is given.
This leads to a formulation
maximizing the empirical fitness,
\begin{align}
\sum_{i, j \in \{1, \cdots, N\}} s_{ij} \sum_{m=1}^M h_m(\mathbf{x}_i) h_m(\mathbf{x}_j),
\label{eqn:semisupersiedhashingobjectivefunction}
\end{align}
where $h_k(\cdot) \in \{1, -1\}$.
It is easily shown that
this objective function~\ref{eqn:semisupersiedhashingobjectivefunction}
is equivalent to
minimizing
$\sum_{i, j \in \{1, \cdots, N\}} s_{ij} \sum_{m=1}^M \frac{(h_m(\mathbf{x}_i) - h_m(\mathbf{x}_j))^2}{2}
=\frac{1}{2}\sum_{i, j \in \{1, \cdots, N\}} s_{ij} \|\mathbf{y}_i - \mathbf{y}_j\|_2^2$.

In addition,
the bit balance requirement (over each hash bit)
is explained as
maximizing the variance over the hash bits.
Assuming the hash function is a sign function,
$h(\mathbf{x}) = \operatorname{sign}(\mathbf{w}^T\mathbf{x})$,
variance maximization is relaxed
as maximizing the variance of the projected data $\mathbf{w}^T\mathbf{x}$.
In summary,
the formulation is given as
\begin{align}
\operatorname{trace}[\mathbf{W}^T\mathbf{X}_l\mathbf{S}\mathbf{X}_l^T\mathbf{W}]
+\eta \operatorname{trace}[\mathbf{W}^T\mathbf{X}\mathbf{X}^T\mathbf{W}],
\end{align}
where $\mathbf{S}$ is the similarity matrix over the labeled data $\mathbf{X}_l$,
$\mathbf{X}$ is the data matrix withe each column corresponding to one data item,
and $\eta$ is a balance variable.

In the case that $\mathbf{W}$
is an orthogonal matrix (the columns are orthogonal to each other,
$\mathbf{W}^T\mathbf{W} = \mathbf{I}$, which is called projection uncorrelation)
(equivalent to the independence requirement in spectral hashing),
it is solved by eigen-decomposition.
The authors present a sequential projection learning algorithm
by embedding $\mathbf{W}^T\mathbf{W} = \mathbf{I}$ into the objective function
as a soft constraint
\begin{align}
&~\operatorname{trace}[\mathbf{W}^T\mathbf{X}_l\mathbf{S}\mathbf{X}_l^T\mathbf{W}]
+\eta \operatorname{trace}[\mathbf{W}^T\mathbf{X}\mathbf{X}^T\mathbf{W}] \nonumber \\
&~+ \rho \|\mathbf{W}^T\mathbf{W} - \mathbf{I}\|_F^2,
\end{align}
where $\rho$ is a tradeoff variable.
An extension of semi-supervised hashing to nonlinear hash functions
is presented in~\cite{WuZCCB13},
where the kernel hash function,
$h(\mathbf{x})
= \operatorname{sign}(\sum_{t=1}^{T} w_{t}<\phi(\mathbf{s}_{t}), \phi(\mathbf{x})> - b_)$, is used.

\subsubsection{LDA hash}

LDA (linear discriminant analysis) hash~\cite{StrechaBBF12} aims to find the binary codes
by minimizing the following objective function,
\begin{align}
\alpha \operatorname{E}
\{\|\mathbf{y}_i - \mathbf{y}_j\|^2 | (i,j) \in \mathcal{P}\}
-
\operatorname{E}
\{\|\mathbf{y}_i - \mathbf{y}_j\|^2 | (i,j) \in \mathcal{N}\},
\end{align}
where $\mathbf{y} = \operatorname{sign}(\mathbf{W}^T\mathbf{x} + \mathbf{b})$,
$\mathcal{P}$ is the set of positive (similar) pairs,
and $\mathcal{N}$ is the set of negative (dissimilar) pairs.

LDA hash
consists of two steps:
(1) finding the projection matrix
that best discriminates
the nearer pairs
from the farther pairs,
which is a form of coding consistency,
and (2) finding the threshold
to generate binary hash codes.
The first step relaxes the problem,
by removing the $\operatorname{sign}$
and minimizes a related function,
\begin{align}
&~\alpha \operatorname{E}
\{\|\mathbf{W}^T\mathbf{x}_i - \mathbf{W}^T\mathbf{x}_j\|^2 | (i,j) \in \mathcal{P}\} \nonumber  \\
&~-
\operatorname{E}
\{\|\mathbf{W}^T\mathbf{x}_i - \mathbf{W}^T\mathbf{x}_j\|^2 | (i,j) \in \mathcal{N}\}.
\end{align}
This formulation is then transformed
to an equivalent form,
\begin{align}
\alpha\operatorname{trace}\{\mathbf{W}^T\boldsymbol{\Sigma}_p \mathbf{W}\}
- \operatorname{trace}\{\mathbf{W}^T\boldsymbol{\Sigma}_n \mathbf{W}\},
\end{align}
where $\boldsymbol{\Sigma}_p = \operatorname{E}
\{(\mathbf{x}_i - \mathbf{x}_j)(\mathbf{x}_i - \mathbf{x}_j)^T |
(i,j) \in \mathcal{P}\}$
and
$\boldsymbol{\Sigma}_n = \operatorname{E}
\{(\mathbf{x}_i - \mathbf{x}_j)(\mathbf{x}_i - \mathbf{x}_j)^T |
(i,j) \in \mathcal{N}\}$.
There are two solutions given in~\cite{StrechaBBF12}:
minimizing $\operatorname{trace}\{\mathbf{W}^T\boldsymbol{\Sigma}_p
\boldsymbol{\Sigma}_n^{-1}\mathbf{W}\}$,
which does not need to specify $\alpha$,
and minimizing
$\operatorname{trace}\{\mathbf{W}^T (\alpha\boldsymbol{\Sigma}_p - \boldsymbol{\Sigma}_n)\}$.

The second step
aims to find the threshold
by minimizing
\begin{align}
\alpha\operatorname{E}\{\operatorname{sign}\{\mathbf{W}^T\mathbf{x}_i - \mathbf{b}\}
- \operatorname{sign}\{\mathbf{W}^T\mathbf{x}_j - \mathbf{b}\} | (i,j) \in \mathcal{P}\} \\
-
\operatorname{E}\{\operatorname{sign}\{\mathbf{W}^T\mathbf{x}_i - \mathbf{b}\}
- \operatorname{sign}\{\mathbf{W}^T\mathbf{x}_j - \mathbf{b}\} | (i,j) \in \mathcal{N}\},
\end{align}
which is then decomposed into $K$ subproblems
each of which finds $b_k$ for each hash function
$\mathbf{w}_k^T\mathbf{x} - b_k$.
The subproblem can be exactly solved
using simple $1$D search.


\subsubsection{Topology preserving hashing}
Topology preserving hashing~\cite{ZhangZTGLT13} formulates the hashing problem
by considering two forms of coding consistency:
preserving the neighborhood ranking
and preserving the data topology.

The first coding consistency form is presented as a maximization problem,
\begin{align}
&~\frac{1}{2}\sum_{i,j, s, t}\operatorname{sign}(d^o_{i,j} - d^o_{s,t})\operatorname{sign}(d^h_{i,j} - d^h_{s,t})\\
\approx&~ \frac{1}{2}\sum_{i,j, s, t}(d^o_{i,j} - d^o_{s,t})(d^h_{i,j} - d^h_{s,t})
\end{align}
where $d^o$ and $d^h$
are the distances
in the original space and the Hamming space.
This ranking preserving formulation,
based on the rearrangement inequality,
is transformed to
\begin{align}
&~\frac{1}{2}\sum_{i,j}d^o_{i,j}d^h_{i,j}\\
~=&~ \frac{1}{2}\sum_{i,j}d^o_{i,j}\|\mathbf{y}_i - \mathbf{y}_j\|_2^2 \\
~=&~ \operatorname{trace}(\mathbf{Y}\mathbf{L}_t\mathbf{Y}^T),
\end{align}
where $\mathbf{L}_t = \mathbf{D}_t - \mathbf{S}_t$,
$\mathbf{D}_t = \operatorname{diag}(\mathbf{S}_t\mathbf{1})$
and $s_t(i,j) = f(d^o_{ij})$ with $f(\cdot)$ is monotonically non-decreasing.

Data topology preserving is formulated in a way
similar to spectral hashing,
by minimizing the following function
\begin{align}
~&~\frac{1}{2} \sum_{ij}s_{ij} \|\mathbf{y}_i - \mathbf{y}_j\|_2^2 \\
=&~\operatorname{trace}(\mathbf{Y}\mathbf{L}_s\mathbf{Y}^T),
\end{align}
where
$\mathbf{L}_s = \mathbf{D}_s - \mathbf{S}_s$,
$\mathbf{D}_s = \operatorname{diag}(\mathbf{S}_s\mathbf{1})$,
and $s_s(i,j) $ is the similarity between $\mathbf{x}_i$ and $\mathbf{x}_j$
in the original space.

Assume the hash function is in the form of $\operatorname{sign}(\mathbf{W}^T\mathbf{x})$
(the following formulation can also be extended to the kernel hash function),
the overall formulation, by a relaxation step $\operatorname{sign}(\mathbf{W}^T\mathbf{x}) \approx \mathbf{W}^T\mathbf{x}$,
is given as follows,
\begin{align}
\max~&~\frac{\operatorname{trace}(\mathbf{W}^T\mathbf{X}(\mathbf{L}_t + \alpha\mathbf{I})\mathbf{X}^T\mathbf{W})}
{\operatorname{trace}(\mathbf{W}^T\mathbf{X}\mathbf{L}_s\mathbf{X}^T\mathbf{W})},
\end{align}
where $\alpha\mathbf{I}$ introduces a regularization term,
$\operatorname{trace}(\mathbf{W}^T\mathbf{X}\mathbf{X}^T\mathbf{W})$,
similar to the bit balance condition in semi-supervised hashing~\cite{WangKC10a, WangKC10b, WangKC12}.

\subsubsection{Hashing with graphs}

The key ideas of hashing with graphs~\cite{LiuWKC11}
consist of using the anchor graph to approximate the neighborhood graph,
(accordingly using the graph Laplacian over the anchor graph
to approximate the graph Laplacian of the original graph)
for fast computing the eigenvectors
and using a hierarchical hashing to address the boundary issue
for which the points around the hash plane are assigned different hash bits.
The first idea aims to solve the same problem in spectral hashing~\cite{WeissTF08},
present an approximate solution using the anchor graph
rather than the PCA-based solution with the assumption
that the data points are uniformly distributed.
The second idea breaks the independence constraint
over hash bits.


Compressed hashing~\cite{LinJCYL13}
borrows the idea about anchor graph in~\cite{LiuWKC11}
uses the anchors
to generate a sparse representation of
data items
by computing the kernels with the nearest anchors and
normalizing it so that the summation is $1$.
Then it uses $M$ random projections
and the median of the projections of the sparse projections
along each random projection as the bias
to generate the hash functions.

\subsection{Similarity Alignment Hashing}
\label{sec:sah}
Similarity alignment hashing
is a category of hashing algorithms
that directly compare the similarities (distances)
computed from the input space
and the coding space.
In addition, the approach aligning the distance distribution
is also discussed in this section.
Other algorithms, such as quantization,
can also be interpreted
as similarity alignment,
and for clarity, are described in separate paragraphs.

\subsubsection{Binary reconstructive embedding}
The key idea of binary reconstructive embedding~\cite{KulisD09}
is to learn the hash codes
such that the difference between the Euclidean distance
in the input space
and the Hamming distance
in the hash codes
is minimized.
The objective function is formulated as follows,
\begin{align}
\min\sum_{(i,j) \in \mathcal{N}} (\frac{1}{2} \|\mathbf{x}_i - \mathbf{x}_j\|_f^2 - \frac{1}{M} \|\mathbf{y}_i - \mathbf{y}_j\|_2^2)^2.\label{eqn:objectiveFunctionForBRE}
\end{align}
The set $\mathcal{N}$ is composed of point pairs,
which includes
both the nearest neighbors
and other pairs.

The hash function is parameterized as:
\begin{align}
y_{nm} = h_m(\mathbf{x})
= \operatorname{sign}(\sum_{t=1}^{T_m} w_{mt} K(\mathbf{s}_{mt}, \mathbf{x})),
\end{align}
where $\{\mathbf{s}_{mt}\}_{t=1}^{T_m}$ are sampled data items
forming the hashing function $h_m(\cdot) \in \{h_1(\cdot), \cdots, h_M(\cdot)\}$,
$K(\cdot,\cdot)$ is a kernel function,
and $\{w_{mt}\}$ are the weights
to be learnt.

Instead of relaxing the $\operatorname{sign}$ function
to a continuous function,
an alternative optimization scheme is presented in~\cite{KulisD09}:
fixing all but one weight $w_{mt}$
and optimizing the problem~\ref{eqn:objectiveFunctionForBRE}
with respect to $w_{mt}$.
It is shown that
an exact, optimal update to this weight $w_{mt}$
(fixing all the other weights)
can be achieved in time $O(N\log N + n|\mathcal{N}|)$.

%

\subsubsection{Supervised hashing with kernels}
The idea of supervised hashing with kernels~\cite{LiuWJJC12}
consists of two aspects:
(1) using the kernels to form the hash functions,
which is similar to binary reconstructive embedding~\cite{KulisD09},
and (2) minimizing the differences
between the Hamming affinity over the hash codes
and the similarity over the data items,
which has two types, similar ($s=1$) or dissimilar ($s=-1$)
e.g., given by the Euclidean distance or the labeling information.

The hash function is given as follows,
\begin{align}
y_{nm} = h_m(\mathbf{x}_n)
= \operatorname{sign}(\sum_{t=1}^{T_m} w_{mt} K(\mathbf{s}_{mt}, \mathbf{x}) + b),
\end{align}
where $b$ is the bias. The objective function is given as the following,
\begin{align}
\min\sum_{(i,j) \in \mathcal{L}} (s_{ij}  - \operatorname{affinity}(\mathbf{y}_i, \mathbf{y}_j))^2,\label{eqn:objectiveFunctionForHashwithKernelsWeiLiu}
\end{align}
where $\mathcal{L}$ is the set of labeled pairs,
$\operatorname{affinity}(\mathbf{y}_i, \mathbf{y}_j) = M - \|\mathbf{y}_i - \mathbf{y}_j\|_1$
is the Hamming affinity,
and $\mathbf{y} \in \{1, -1\}^M$.

Kernel reconstructive hashing~\cite{YangBZRZC14}
extends this technique
using a normalized Gaussian kernel similarity.

\subsubsection{Spec hashing}
The idea of spec hashing~\cite{LinRY10}
is to view each pair of data items as a sample
and their (normalized) similarity
as the probability,
and to find the hash functions
so that
the probability distributions
from the input space and the Hamming space
are well aligned.
Let $s^i_{ij}$
be the normalized similarity
($\sum_{ij}s^i_{ij} = 1$) given in the input space,
and $s^h_{ij}$
be the normalized similarity
computed in the Hamming space,
$s^h_{ij} = \frac{1}{Z} \exp{(-\lambda \operatorname{dist}_h(i,j))}$,
where $Z$ is a normalization variable
$Z = \sum_{ij} \exp{(-\lambda \operatorname{dist}_h(i,j))}$.
Then,
the objective function is given as follows,
\begin{align}
\min~&~\operatorname{KL} (\{s^i_{ij}\} || \{s^h_{ij}\}) \nonumber \\
&~= - \sum_{ij} \lambda s^i_{ij} \log s^h_{ij}\\
&~= \lambda \sum_{ij} s^u_{ij} \operatorname{dist}_h(i,j) + \log \sum_{ij} \exp{(-\lambda \operatorname{dist}_h(i,j))}.
\end{align}

Supervised binary hash code learning~\cite{Fan13}
presents a supervised binary hash code learning algorithm
using Jensen Shannon Divergence
which is derived from
minimizing an upper bound
of the probability of Bayes decision errors.


\subsubsection{Bilinear hyperplane hashing}
Bilinear hyperplane hashing~\cite{LiuWMKC12}
transforms the database vector
(the normal of the query hyperplane)
into a high-dimensional vector,
\begin{align}
\bar{\mathbf{a}}
= \operatorname{vec}(\mathbf{a} \mathbf{a}^T)
[a_1^2, a_1a_2, \cdots, a_1a_d, a_2a_1, a_2^2,a_2a_3, \cdots, a_d^2].
\end{align}

The bilinear hyperplane hashing family is defined
as follows,
\begin{equation}
h(\mathbf{z}) = \left\{ \begin{array}{l l}
     \operatorname{sign}(\mathbf{u}^T\mathbf{z}\mathbf{z}^T\mathbf{v})  &  \text{if $\mathbf{z}$ is a database vector}\\
     \operatorname{sign}(-\mathbf{u}^T\mathbf{z}\mathbf{z}^T\mathbf{v})  & \text{if $\mathbf{z}$ is a hyperplane normal.}
   \end{array} \right.
\end{equation}
Here $\mathbf{u}$ and $\mathbf{v}$
are sampled independently from a standard Gaussian distribution.
It is shown to be $r, r(1+\epsilon), \frac{1}{2} - \frac{2r}{\pi^2}, \frac{1}{2} - \frac{2r(1+\epsilon)}{\pi^2}$-sensitive
to the angle distance $d_{\theta}(\mathbf{x}, \mathbf{n}) = (\theta_{\mathbf{x}, \mathbf{n}} - \frac{\pi}{2})^2$,
where $r, \epsilon > 0$.

Rather than randomly drawn,
$\mathbf{u}$ and $\mathbf{u}$ can be also learnt according to the similarity information.
A formulation is given in~\cite{LiuWMKC12} as the below,
\begin{align}
\min_{\{\mathbf{u}_k, \mathbf{v}_k\}_{k=1}^K}
\|\frac{1}{K} \mathbf{Y}^T\mathbf{Y} - \mathbf{S}\|,
\end{align}
where $\mathbf{Y} = [\mathbf{y}_1, \mathbf{y}_2, \cdots, \mathbf{y}_N]$
and $\mathbf{S}$ is the similarity matrix,
\begin{equation}
s_{ij} = \left\{ \begin{array}{l l}
     1  & \quad \text{if $\cos(\theta_{\mathbf{x}_i, \mathbf{x}_j}) \geqslant t_1$}\\
     -1  & \quad \text{if $\cos(\theta_{\mathbf{x}_i, \mathbf{x}_j}) \leqslant t_2$}\\
     2|\cos(\theta_{\mathbf{x}_i, \mathbf{x}_j})| - 1 & \text{otherwise,}
   \end{array} \right.
\end{equation}

The above problem is solved
by relaxing $\operatorname{sign}$
with the sigmoid-shaped function
and finding the solution with the gradient descent algorithm.

\subsection{Order Preserving Hashing}
\label{sec:oph}
This section reviews the category of hashing algorithms
that depend on
various forms of
maximizing the alignment between the orders
of the reference data items
computed from the input space and the coding space.

\subsubsection{Minimal loss hashing}
The key point of minimal loss hashing~\cite{NorouziF11}
is to use a hinge-like loss function
to assign penalties
for similar (or dissimilar) points
when they are too far apart (or too close).
The formulation is given as follows,
\begin{align}
\min &~\sum_{(i,j) \in \mathcal{L}} I[s_{ij} = 1] \max(\|\mathbf{y}_i - \mathbf{y}_i\|_1 - \rho + 1, 0) \nonumber \\
&~ + I[s_{ij} = 0] \lambda \max(\rho - \|\mathbf{y}_i - \mathbf{y}_i\|_1 + 1, 0),
\end{align}
where $\rho$ is a hyper-parameter
and is uses as a threshold
in the Hamming space
that differentiates neighbors
from non-neighbors,
$\lambda$ is also a hyper-parameter
that controls the ratio
of the slopes
for the penalties
incurred for similar (or dissimilar) points.
Both the two hyper-parameters
are selected using the validation set.

Minimal loss hashing~\cite{NorouziF11}
solves the problem
by building the convex-concave upper bound
of the above objective function
and optimizing it using the perceptron-like learning procedure.

%
%

\subsubsection{Rank order loss}
The idea of order preserving hashing~\cite{WangWYL13}
is to learn hash functions by maximizing
the alignment between the similarity orders computed
from the original space and the ones in the Hamming
space.
To formulate the problem,
given a data point $\mathbf{x}_n$,
the database points $\mathcal{X}$
are divided into $M$ categories,
$(\mathcal{C}^e_{n0},~\mathcal{C}^e_{n1},~\cdots,~\mathcal{C}^e_{nM})$
and
$(\mathcal{C}^h_{n0},~\mathcal{C}^h_{n1},~\cdots,~\mathcal{C}^h_{nM})$,
using the distance in the original space
and the distance in the Hamming space, respectively.
The objective function maximizing the alignment between the two categories
is given as follows,
\begin{align}
&~L(\mathbf{h}(\cdot); \mathcal{X})\\
=&~\sum\nolimits_{n=1}^N L(\mathbf{h}(\cdot); \mathbf{x}_n) \\
=&~\sum\nolimits_{n=1}^N \sum\nolimits_{m=0}^{M-1} L(\mathbf{h}(\cdot); \mathbf{x}_n, m) \\
=&~\sum\nolimits_{n=1}^N \sum\nolimits_{m=0}^{M-1} (|\mathcal{N}^e_{nm} - \mathcal{N}^h_{nm}| +
\lambda |\mathcal{N}^h_{nm} - \mathcal{N}^e_{nm}|),
\end{align}
where
$\mathcal{N}^e_{nm} = \cup_{j = 0}^m{\mathcal{C}^e_{nj}}$
and $\mathcal{N}^h_{nm} = \cup_{j = 0}^m{\mathcal{C}^h_{nj}}$.

Given the compound hash function defined as below,
\begin{align}
\mathbf{h}(\mathbf{x})
= &~\operatorname{sign}(\mathbf{W}^T\mathbf{x} + \mathbf{b})\label{eqn:hashfunction}\\
= &~[\operatorname{sign}(\mathbf{w}_1^T\mathbf{x} + b_1)~\cdots~\operatorname{sign}(\mathbf{w}_m^T\mathbf{x} + b_m)]^T\notag,
\end{align}
the loss is transformed to:
\begin{align}
& ~L(\mathbf{W}; \mathbf{x}_n, i) \nonumber\\
=&~ \sum\nolimits_{\mathbf{x}' \in \mathcal{N}^e_{ni}} \operatorname{sign} (\|\mathbf{h}(\mathbf{x}_n) - \mathbf{h}(\mathbf{x}')\|^2_2 -i) \nonumber \\
 &~+ \lambda\sum\nolimits_{\mathbf{x}' \notin \mathcal{N}^e_{ni}} \operatorname{sign} (i + 1 - \|\mathbf{h}(\mathbf{x}_n) - \mathbf{h}(\mathbf{x}')\|^2_2).
\label{eqn:nonsmoothformulation}
\end{align}
This problem is solved by
dropping the $\operatorname{sign}$ function
and using the quadratic penalty algorithm~\cite{WangWYL13}.

\subsubsection{Triplet loss hashing}
Triplet loss hashing~\cite{NorouziFS12} formulates the hashing problem
by preserving the relative similarity defined over
triplets of items,
$(\mathbf{x}, \mathbf{x}^+, \mathbf{x}^-)$,
where the pair $(\mathbf{x}, \mathbf{x}^+)$
is more similar than
the pair $(\mathbf{x}, \mathbf{x}^-)$.
The triplet loss is defined as
\begin{align}
\ell_{\text{triplet}} (\mathbf{y}, \mathbf{y}^+, \mathbf{y}^-)
= \max (\|\mathbf{y} - \mathbf{y}^+\|_1 - \|\mathbf{y} - \mathbf{y}^-\|_1 + 1, 0).
\end{align}

Suppose the compound hash function is defined as
$\mathbf{h}(\mathbf{x};\mathbf{W})$,
the objective function is given as follows,
\begin{align}
\sum_{(\mathbf{x}, \mathbf{x}^+, \mathbf{x}^-) \in \mathcal{D}} &
\ell_{\text{triplet}} (\mathbf{h}(\mathbf{x}; \mathbf{W}), \mathbf{h}(\mathbf{x}^+; \mathbf{W}), \mathbf{h}(\mathbf{x}^-; \mathbf{W})) \nonumber \\
&+ \frac{\lambda}{2} \operatorname{trace}{(\mathbf{W}^T\mathbf{W})}.
\end{align}
The problem is optimized
using the algorithm similar to minimal loss hashing~\cite{NorouziF11}.
The extension to asymmetric Hamming distance is also discussed in~\cite{NorouziF11}.

\subsubsection{Listwise supervision hashing}
Similar to~\cite{NorouziF11},
listwise supervision hashing~\cite{WangLSJ13} also uses triplets of items to approximate the listwise loss.
The formulation is based on a triplet tensor $\mathbf{S}$ defined as follows,
\begin{equation}
s(i; j, k)
= \left\{ \begin{array}{l l}
     1 & \quad \text{if $\operatorname{sim}(\mathbf{x}_i; \mathbf{i}) >  \operatorname{sim}(\mathbf{x}_i; \mathbf{j})$}\\
     -1 & \quad \text{if $\operatorname{sim}(\mathbf{x}_i; \mathbf{i}) <  \operatorname{sim}(\mathbf{x}_i; \mathbf{j})$}\\
     0 & \quad \text{if $\operatorname{sim}(\mathbf{x}_i; \mathbf{i}) =  \operatorname{sim}(\mathbf{x}_i; \mathbf{j})$}.\\
   \end{array} \right.
\end{equation}

The goal is to minimize the following objective function,
\begin{align}
-\sum_{i, j, k} \mathbf{h}(\mathbf{x}_i)^T(\mathbf{h}(\mathbf{x}_j) - \mathbf{h}(\mathbf{x}_k))s_{ijk},
\end{align}
where is solved by dropping the $\operatorname{sign}$ operator in
$\mathbf{h}(\mathbf{x};\mathbf{W}) = \operatorname{sign}(\mathbf{W}^T\mathbf{x})$.

\subsubsection{Similarity sensitive coding}
Similarity sensitive coding (SSC)~\cite{Shakhnarovich05}
aims to learn an embedding,
which can be called weighted Hamming embedding:
$\mathbf{h}(\mathbf{x}) = [\alpha_1 h(\mathbf{x}_1)~\alpha_2 h(\mathbf{x}_2)~\cdots~\alpha_M h(\mathbf{x}_M)]$
that is faithful to a task-specific similarity.
An example algorithm,
boosted SSC,
uses adaboost to learn a classifier.
The output of each weak learner
on an input item
is a binary code,
and the outputs
of all the weak learners
are aggregated as the hash code.
The weight of each weak learner
forms the weight in the embedding,
and is used to compute the weighted Hamming distance.
Parameter sensitive hashing~\cite{ShakhnarovichVD03}
is a simplified version of SSC with the standard LSH search procedure
instead of the linear scan with weighted Hamming distance
and uses decision stumps to form hash functions
with threshold optimally decided
according to the information
of similar pairs,
dissimilar pairs and
pairs with undefined similarities.
The forgiving hashing approach~\cite{BalujaC07,BalujaC08, BalujaC10}
extends parameter sensitive hashing
and does not explicitly create dissimilar pairs,
but instead relies on the maximum entropy constraint
to provide that separation.

A column generation algorithm, which can be used to solve adaboost, is presented to
simultaneously learn the weights and hash functions~\cite{LiLSHD13},
with the following objective function
\begin{align}
\min_{\boldsymbol{\alpha} , \boldsymbol{\zeta}} ~&~
\sum_{i=1}^N \zeta_i + C \| \boldsymbol{\alpha} \|_p \\
\operatorname{s.t.}~&~
\boldsymbol{\alpha} \geqslant \mathbf{0}, \boldsymbol{\zeta} \geq \mathbf{0},\\
~&~
d_h(\mathbf{x}_i, \mathbf{x}_i^-) - d_h(\mathbf{x}_i, \mathbf{x}_i^+) \geqslant 1 - \zeta_i \forall i.
\end{align}
Here $\|\cdot\|_l$ is a $\ell_p$ norm, e.g., $l=1, 2, \infty$.

\subsection{Regularized Space Partitioning}
\label{sec:rsp}
Almost all hashing algorithms
can be interpreted
from the view of partitioning the space.
In this section,
we review the category of hashing algorithms
that focus on pursuiting effective space partitioning
without explicitly evaluating the distance
in the coding space.

\subsubsection{Complementary projection hashing}
Complementary projection hashing~\cite{JinHLZLCL13} computes
the $m$th hash function
according to the previously computed $(m-1)$ hash functions,
using a way similar to complementary hashing~\cite{XuWLZLY11},
checking the distance of the point to the previous $(m-1)$ partition planes.
The penalty weight for $\mathbf{x}_n$ when learning the $m$th hash function
is given as
\begin{align}
u_n^m =
1 + \sum_{j=1}^{m-1} H(\epsilon - |\mathbf{w}_m^T\mathbf{x}_n + b|),
\end{align}
where $H(\cdot) = \frac{1}{2}(1+ \operatorname{sign}(\cdot))$ is a unit function.

Besides, it generalizes the bit balance condition,
for each hit, half of points are mapped to $-1$
and the rest mapped to $1$,
and introduces a pair-wise bit balance condition
to approximate the coding balance condition,
i.e.
every two hyperplanes spit the space into four subspaces,
and each subspace contains $N/4$ data points.
The condition is guaranteed by
\begin{align}
\sum_{n=1}^N h_1(\mathbf{x}_n) &~= 0, \\
\sum_{n=1}^N h_2(\mathbf{x}_n) &~= 0, \\
\sum_{n=1}^N h_1(\mathbf{x}_n) h_2(\mathbf{x}_n) &~= 0.
\end{align}
The whole formulation for updating the $m$th hash function
is written as the following
\begin{align}
\min ~&~\sum_{n=1}^N u_n^m H(\epsilon - |\mathbf{w}_m^T\mathbf{x}_n + b|)
+ \alpha ( (\sum_{n=1}^N h_m(\mathbf{x}_n) )^2 \nonumber \\
&~ + \sum_{j=1}^{m-1} (\sum_{n=1}^N h_j(\mathbf{x}_n) h_m(\mathbf{x}_n))^2,
\end{align}
where $h_m(\mathbf{x}) = \operatorname{(\mathbf{w}_m^T\mathbf{x} + b)}$.

The paper~\cite{JinHLZLCL13} also extends the linear hash function
the kernel function,
and presents the gradient descent algorithm
to optimize the continuous-relaxed objective function
which is formed by dropping the $\operatorname{sign}$ function.


\subsubsection{Label-regularized maximum margin hashing}
The idea of label-regularized maximum margin hashing~\cite{MuSY10} is to use the side information
to find the hash function
with the maximum margin criterion.
Specifically,
the hash function is computed
so that ideally one pair of similar points are mapped to the same hash bit
and one pair of dissimilar points are mapped to different hash bits.
Let $\mathcal{P}$
be a set of pairs $\{(i, j)\}$ labeled
to be similar.
The formulation is given as follows,
\begin{align}
\min_{\{y_i\}, \mathbf{w}, b, \{\xi_i\}, \{\zeta\}}
~&~\|\mathbf{w}\|_2^2 + \frac{\lambda_1}{N}\sum_{n=1}^N \xi_n + \frac{\lambda_2}{N} \sum_{(i, j) \in \mathcal{S}} \zeta_{ij} \\
\operatorname{s.t.}~&~
y_i(\mathbf{w}^T\mathbf{x}_i + b) + \xi_i \geqslant 1, \xi_i \geqslant 0, \forall i, \\
~&~
y_iy_j + \zeta_{ij} \geqslant 0. \forall (i,j) \in \mathcal{P}, \\
~&~ -l \leqslant \mathbf{w}^T\mathbf{x}_i + b \leqslant l.
\end{align}
Here,
$\|\mathbf{w}\|_2^2$ corresponds to the maximum margin criterion.
The second constraint comes from the side information
for similar pairs,
and its extension to dissimilar pairs is straightforward.
The last constraint comes from the bit balance constraint,
half of data items mapped to $-1$ or $1$.

Similar to the BRE,
the hash function is defined as
$h(\mathbf{x})
= \operatorname{sign}(\sum_{t=1}^{T} v_{t}<\phi(\mathbf{s}_{t}), \phi(\mathbf{x})> - b_)$,
which means that
$w = \sum_{t=1}^{T} v_{t}\phi(\mathbf{s}_{t})$.
This definition reduces the optimization cost.
Constrained-concave-convex-procedure (CCCP)
and cutting plane are used for the optimization.

%

\subsubsection{Random maximum margin hashing}
Random maximum margin hashing~\cite{JolyB11} learns
a hash function with the maximum margin criterion,
where the positive and negative labels
are randomly generated,
by randomly sampling $N$ data items
and randomly labeling half of the items
with $-1$
and the other half with $1$.
The formulation is a standard SVM formulation
that is equivalent to the following form,
\begin{align}
\max \frac{1}{\|\mathbf{w}\|_2}
\min[\min_{i=1}^{\frac{N'}{2}}(\mathbf{w}^T\mathbf{x}_i^+ + b), \min_{i=1}^{\frac{N'}{2}}(-\mathbf{w}^T\mathbf{x}_i^- - b)],
\end{align}
where $\{\mathbf{x}_i^+\}$ are the positive samples
and $\{\mathbf{x}_i^-\}$ are the negative samples.
Using the kernel trick,
the hash function can be a kernel-based function,
$h(\mathbf{x} = \operatorname{sign}(\sum_{i=1}^v\alpha_i<\phi(\mathbf{x}), \phi(\mathbf{s})>) + b)$,
where $\{\mathbf{s}\}$ are the selected $v$ support vectors.


\subsubsection{Spherical hashing}
The basic idea of spherical hashing~\cite{HeoLHCY12}
is to use a hypersphere
to formulate a spherical hash function,
\begin{equation}
h(\mathbf{x}) = \left\{ \begin{array}{l l}
     +1 & \quad \text{if $d(\mathbf{p}, \mathbf{x}) \leqslant t$}\\
     0 & \quad \text{otherwise.}
   \end{array} \right.
\end{equation}
The compound hash function
consists of $K$ spherical functions,
depending on
$K$ pivots $\{\mathbf{p}_1, \cdots, \mathbf{p}_K\}$
and $K$ thresholds $\{t_1, \cdots, t_K\}$.
Given two hash codes,
$\mathbf{y}_1$ and $\mathbf{y}_2$,
the distance is computed as
\begin{align}
\frac{\|\mathbf{y}_1 - \mathbf{y}_2\|_1}{ \mathbf{y}_1^T  \mathbf{y}_2},
\end{align}
where $\|\mathbf{y}_1 - \mathbf{y}_2\|_1$ is similar to the Hamming distance,
i.e.,
the frequency
that both the two points lie inside (or outside) the hypersphere,
and $ \mathbf{y}_1^T  \mathbf{y}_2$ is equivalent to
the number of common $1$ bits
between two binary codes,
i.e.,
the frequency
that both the two points lie inside the hypersphere.

The paper~\cite{HeoLHCY12}
proposes an iterative optimization algorithm
to learn $K$ pivots and thresholds
such that
it satisfies a pairwise bit balanced condition:
$$\|\{\mathbf{x}|h_k(\mathbf{x}) = 1\}\|
= \|\{\mathbf{x}|h_k(\mathbf{x}) = 0\}\|,$$
and
$$\|\{\mathbf{x}|h_i(\mathbf{x}) = b_1, h_j(\mathbf{x}) = b_2\}\|
= \frac{1}{4} \|\mathcal{X}\|, b_1, b_2 \in \{0, 1\}.$$


\subsubsection{Density sensitive hashing}
The idea of density sensitive hashing~\cite{LinCL12}
is to exploit the clustering results
to generate a set of candidate hash functions
and to select the hash functions
which can split the data most equally.
First, the $k$-means algorithm is
run over the data set,
yielding $K$ clusters
with centers being $\{\boldsymbol{\mu}_1, \boldsymbol{\mu}_2, \cdots, \boldsymbol{\mu}_K\}$.
Second, a hash function is defined
over two clusters $(\boldsymbol{\mu}_i, \boldsymbol{\mu}_j)$
if the center is one of the $r$ nearest neighbors of the other,
$h(\mathbf{x}) = \operatorname{sign} (\mathbf{w}^T \mathbf{x} - b)$,
where $\mathbf{w} = \boldsymbol{\mu}_i - \boldsymbol{\mu}_j$
and $b = \frac{1}{2}(\boldsymbol{\mu}_i + \boldsymbol{\mu}_j)^T (\boldsymbol{\mu}_i - \boldsymbol{\mu}_j)$.
The third step
aims to evaluate if
the hash function $(\mathbf{w}_m, b_m)$ can split the data most equally,
which is evaluated by the entropy,
$-P_{m0}\log P_{m0} - -P_{m1}\log P_{m1}$,
where $P_{m0} =\frac{n_0}{n}$
and $P_{m1} =1 - P_{m0}$.
$n$ is the number of the data points,
and $n_0$ is the number of the data points lying one partition
formed by the hyperplane of the corresponding hash function.
Lastly,
$L$ hash functions with the greatest entropy scores are selected
to form the compound hash function.

\subsection{Hashing with Weighted Hamming Distance}
\label{sec:whd}
This section presents the hashing algorithms
which evaluates the distance in the coding space
using the query-dependent and query-independent weighted Hamming distance scheme.

\subsubsection{Multi-dimensional spectral hashing}
Multi-dimensional spectral hashing~\cite{WeissFT12} seeks
hash codes
such that
the weighted Hamming affinity is equal to the original affinity,
\begin{align}
\min ~&~ \sum_{(i,j) \in \mathcal{N}} (w_{ij} - \mathbf{y}_i^T\boldsymbol{\Lambda}\mathbf{y}_j)^2 = \|\mathbf{W} - \mathbf{Y}^T \boldsymbol{\Lambda} \mathbf{Y}\|_F^2,
\label{eqn:objectiveFunctionForMultidimensionalSpectralHashing}
\end{align}
where $\boldsymbol{\Lambda}$ is a diagonal matrix,
and both $\boldsymbol{\Lambda}$ and hash codes $\{\mathbf{y}_i\}$ are needed to be optimized.

The algorithm for solving the problem~\ref{eqn:objectiveFunctionForMultidimensionalSpectralHashing}
to compute hash codes
is exactly the same to that given in~\cite{WeissTF08}.
Differently,
the affinity over hash codes
for multi-dimensional spectral hashing
is the weighted Hamming affinity
rather than the ordinary (isotropically weighted) Hamming affinity.
Let $(d, l)$
correspond to the index of
one selected eigenfunction for computing the hash bit,
the $l$ eigenfunction
along the PC direction $d$,
$\mathcal{I} = \{(d, l)\}$
be the set of the indices of all the selected eigenfunctions.
The weighted Hamming affinity using pure eigenfunctions
along (PC) dimension $d$
is computed as
\begin{align}
\operatorname{affinity}_d(i, j)
= \sum_{(d,l) \in \mathcal{I}}\lambda_{dl}
\operatorname{sign}(\phi_{dl}(x_{id}))
\operatorname{sign}(\phi_{dl}(x_{jd})),
\end{align}
where $x_{id}$ is the projection of $\mathbf{x}_i$
along dimension $d$,
$\phi_{dl}(\cdot)$
is the $l$th eigenfunction
along dimension $d$,
$\lambda_{dl}$ is the corresponding eigenvalue.
The weighted Hamming affinity using all the hash codes
is then computed as follows,
\begin{align}
\operatorname{affinity}(\mathbf{y}_i, \mathbf{y}_j)
= \prod_d(1+ \operatorname{affinity}_d(i, j)) - 1.
\end{align}
The computation can be accelerated
using lookup tables.

%
%

\subsubsection{Weighted hashing}
Weighted hashing~\cite{WangZS13}
uses the weighted Hamming distance to evaluate the distance
between hash codes,
$\|\boldsymbol{\alpha}^T(\mathbf{y}_i - \mathbf{y}_j)\|^2_2$.
It optimizes the following problem,
\begin{align}
\min~&~\operatorname{trace}(\operatorname{diag}(\boldsymbol{\alpha}) \mathbf{Y}\mathbf{L}\mathbf{Y}^T)
+ \lambda \|\frac{1}{n} \mathbf{Y}\mathbf{Y}^T - \mathbf{I} \|_F^2 \\
\operatorname{s.t.}~&~
\mathbf{Y} \in \{-1, 1\}^{M\times N}, \mathbf{Y}^T\mathbf{1} = \mathbf{0}\\
~&~
\|\boldsymbol{\alpha}\|_1 = 1 \\
~&~
\frac{\alpha_1}{\operatorname{var}(y_1)} =
\frac{\alpha_2}{\operatorname{var}(y_2)} =
\cdots
=
\frac{\alpha_M}{\operatorname{var}(y_M)},
\end{align}
where $\mathbf{L} = \mathbf{D} - \mathbf{S}$ is the Laplacian matrix.
The formulation is essentially similar to spectral hashing~\cite{WeissTF08},
and the difference lies in including the weights for weighed Hamming distance.

The above problem is solved
by discarding the first constraint
and then binarizing $\mathbf{y}$
at the $M$ medians.
The hash function $\operatorname{\mathbf{w}_m^T\mathbf{x} + b}$ is learnt
by mapping the input $\mathbf{x}$ to a hash bit $y_m$.


\subsubsection{Query-adaptive bit weights}
\cite{JiangWC11, JiangWXC13} presents a weighted Hamming distance measure
by learning the weights from the query information.
Specifically,
the approach learns class-specific bit weights
so that the weighted Hamming distance
between the hash codes belong the class and
the center, the mean of those hash codes
is minimized.
The weight for a specific query
is the average weight of
the weights of the classes
that the query most likely belong to
and that are discovered
using the top similar images
(each of which is associated with a semantic label).

\subsubsection{Query-adaptive hashing}
Query adaptive hashing~\cite{LiuYJHZ13}
aims to select the hash bits (thus hash functions forming the hash bits)
according to the query vector (image).
The approach consists of two steps:
offline hash functions $\mathbf{h}(\mathbf{x}) = \operatorname{sign}(\mathbf{W}^T\mathbf{x})$
($\{h_b(\mathbf{x}) = \operatorname{sign}(\mathbf{w}_b^T\mathbf{x})\}$)
and online hash function selection.
The online hash function selection, given the query $\mathbf{q}$, is formulated
as the following,
\begin{align}
\min_{\boldsymbol{\alpha}} \|\mathbf{q} - \mathbf{W}\boldsymbol{\alpha}\|_2^2 + \rho \|\boldsymbol{\alpha}\|_1.
\end{align}
Given the optimal solution $\boldsymbol{\alpha}^*$, $\alpha_i^* = 0$ means the $i$th hash function is not selected,
and the hash function corresponding to the nonzero entries in $\boldsymbol{\alpha}^*$.
A solution based on biased discriminant analysis is given
to find $\mathbf{W}$,
for which more details can be found from~\cite{LiuYJHZ13}.

\subsection{Other Hash Learning Algorithms}
\subsubsection{Semantic hashing}
Semantic hashing~\cite{SalakhutdinovH07, SalakhutdinovH09}
generate the hash codes,
which can be used to reconstruct the input data,
using the deep generative model
(based on the pretraining technique
and the fine-tuning scheme
originally designed for the restricted Boltzmann machines).
This algorithm does not use any similarity information.
The binary codes can be used for finding similarity data
as they can be used to well reconstruct the input data.

\subsubsection{Spline regression hashing}
Spline regression hashing~\cite{LiuWYZH12}
aims to find a global hash function in the kernel form,
$h(\mathbf{x}) = \mathbf{v}^T\phi(\mathbf{x})$,
such that the hash value from the global hash function
is consistent to those from the local hash functions that corresponds to its neighborhood points.
Each data point corresponds to a local hash function
in the form of spline regression,
$h_n(\mathbf{x} = \sum_{i=1}^t \beta_{ni} p_i(\mathbf{x}))
+ \sum_{i=1}^k\alpha_{ni}g_{ni}(\mathbf{x})$,
where $\{p_i(\mathbf{x})\}$
are the set of primitive polynomials
which can span the polynomial space with a degree less than $s$,
$\{g_{ni}(\mathbf{x})\}$ are the green functions,
and $\{\alpha_{ni}\}$ and $\{\beta_{ni\}}$
are the corresponding coefficients.
The whole formulation is given as follows,
\begin{align}
\min_{\mathbf{v}, \{h_i\}, \{\mathbf{y}_n\}}
~&~
\sum_{n=1}^N
(\sum_{\mathbf{x}_i \in \mathcal{N}_n} \|\mathbf{h}_n(\mathbf{x}_i) - \mathbf{y}_i\|_2^2 + \gamma\psi_n(\mathbf{h}_n) \nonumber \\
&~+ \lambda(\sum_{n=1}^N \|\mathbf{h}(\mathbf{x}_n) - \mathbf{y}_n\|_2^2 + \gamma \|\mathbf{v}\|_2^2).
\end{align}

%

\subsubsection{Inductive manifold hashing}
Inductive manifold mashing~\cite{ShenSSHT13}
consists of three steps:
cluster the data items into $K$ clusters,
whose centers are $\{\mathbf{c}_1, \mathbf{c}_2, \cdots, \mathbf{c}_K\}$,
embed the cluster centers into a low-dimensional space,
$\{\mathbf{y}_1, \mathbf{y}_2, \cdots, \mathbf{y}_K\}$,
using existing manifold embedding technologies,
and finally
the hash function is given as follows,
\begin{align}
\mathbf{h}(\mathbf{x})
= \operatorname{sign}
(\frac{\sum_{k=1}^K w(\mathbf{x}, \mathbf{c}_k) \mathbf{y}_k}{\sum_{k=1}^K w(\mathbf{x}, \mathbf{c}_k)}).
\end{align}

\subsubsection{Nonlinear embedding}
The approach introduced in~\cite{HwangHA12}
is an exact nearest neighbor approach,
which relies on a key inequality,
\begin{align}
\|\mathbf{x}_1 - \mathbf{x}_2\|_2^2
\geqslant d ((\mu_1 - \mu_2)^2 + (\sigma_1 - \sigma_2)^2),
\end{align}
where $\mu = \frac{1}{d}\sum_{i=1}^d x_i$ is the mean
of all the entries of the vector $\mathbf{x}$,
and $\sigma = \frac{1}{d}\sum_{i=1}^d (x_i - \mu)^2$
is the standard deviation.
The above inequality is generalized
by dividing the vector into $M$ subvectors,
with the length of each subvector being $d_m$,
and the resulting inequality is formulated as follows,
\begin{align}
\|\mathbf{x}_1 - \mathbf{x}_2\|_2^2
\geqslant \sum_{m=1}^M d_m ((\mu_{1m} - \mu_{2m})^2 + (\sigma_{1m} - \sigma_{2m})^2).
\end{align}

In the search strategy,
before computing the exact Euclidean distance
between the query and the database point,
the lower bound is first computed
and is compared with the current minimal Euclidean distance,
to determine if the exact distance is necessary to be computed.

\subsubsection{Anti-sparse coding}
The idea of anti-sparse coding~\cite{JegouFF12}
is to learn a hash code
so that \emph{non-zero elements in the hash code as many as possible}.
The binarization process is as follows.
First, it solves the following problem,
\begin{align}
\mathbf{z}^* = \arg\min_{\mathbf{z}: \mathbf{W}\mathbf{z} = \mathbf{x}} \|\mathbf{z}\|_{\infty},
\end{align}
where $\|\mathbf{z}\|_{\infty} = \max_{i \in \{1, 2, \cdots, K\}} |z_i|$,
and $\mathbf{W}$ is a projection matrix.
It is proved that
in the optimal solution (minimizing the range of the components),
$K-d+1$ of the components are stuck to
the limit,
i.e., $z_i =  \pm \|\mathbf{z}\|_{\infty}$.
The binary code (of the length $K$)
is computed as $\mathbf{y} = \operatorname{sign}(\mathbf{z})$.

The distance between the query $\mathbf{q}$ and a vector $\mathbf{x}$
can be evaluated based on the similarity in the Hamming space,
$\mathbf{y}_q^{\mathsf{T}}\mathbf{y}_x$
or the asymmetric similarity
$\mathbf{z}_q^{\mathsf{T}}\mathbf{y}_x$.
The nice property is that the anti-sparse code allows,
up to a scaling factor,
the explicit reconstruction of the original vector $\mathbf{x} \propto \mathbf{W}\mathbf{y}$.

\subsubsection{Two-Step Hashing}

The paper~\cite{LinSSH13}
presents a general two-step approach
to learning-based hashing:
learn binary embedding (codes)
and then learn the hash function mapping the input item
to the learnt binary codes.
An instance algorithm~\cite{LinSSHS14}
uses an efficient GraphCut based block search method
for inferring binary codes for large databases
and trains boosted decision trees
fit the binary codes.

Self-taught hashing~\cite{ZhangWCL10b}
optimizes an objective function,
similar to the spectral hashing,
\begin{align}
\min ~&~ \operatorname{trace}(\mathbf{Y}\mathbf{L}\mathbf{Y}^T) \\
\operatorname{s.t.}
~&~ \mathbf{Y}\mathbf{D}\mathbf{Y}^T = \mathbf{I} \\
~&~ \mathbf{Y}\mathbf{D}\mathbf{1} = \mathbf{0},
\end{align}
where $\mathbf{Y}$ is a real-valued matrix, relaxed from the binary matrix),
$\mathbf{L}$ is the Laplacian matrix
and $\mathbf{D}$ is the degree matrix.
The solution is
the $M$ eigenvectors
corresponding to the smallest $M$ eigenvalues
(except the trivial eigenvalue $0$),
$\mathbf{L}\mathbf{v} = \lambda \mathbf{D}\mathbf{v}$.
To get the binary code,
each row of $\mathbf{Y}$ is
thresholded
using the median value of the column.
To form the hash function,
mapping the vector to a single hash bit
is regarded
as a classification problem,
which is solved by linear SVM,
$\operatorname{sign}(\mathbf{w}^T\mathbf{x} + b)$.
The linear SVM is then
regarded as the hash function.

Sparse hashing~\cite{ZhuHCCS13}
also is a two step approach.
The first step learns a sparse nonnegative embedding,
in which the positive embedding
is encodes as $1$
and the zero embedding is encoded as $0$.
The formulation is as follows,
\begin{align}
\sum_{n=1}^N \|\mathbf{x}_n - \mathbf{P}^T\mathbf{z}_n\|_2^2
+ \alpha \sum_{i=1}^N \sum_{j=1}^N s_{ij}\|\mathbf{z}_i - \mathbf{z}_j\|_2^2
+ \lambda \sum_{n=1}^N \|\mathbf{z}_n\|_1,
\end{align}
where $s_{ij} = \exp{(-\frac{\|\mathbf{x}_i - \mathbf{x}_j\|_2^2}{\sigma^2})}$ is the similarity
between $\mathbf{x}_i$ and $\mathbf{x}_j$.

The second step is to learn a linear hash function
for each hash bit,
which is optimized
based on the elastic net estimator
(for the $m$th hash function),
\begin{align}
\min_{\mathbf{w}} \sum_{n=1}^N \|\mathbf{y}_n - \mathbf{w}_m^t\mathbf{x}_n\|_2^2
+ \lambda_1 \|\mathbf{w}_m\|_1
+ \lambda_2 \|\mathbf{w}_m\|_2^2.
\end{align}

Locally linear hashing~\cite{IrieLWC14}
first learns binary codes that preserves the locally linear structures
and then introduces a locally linear extension algorithm for out-of-sample extension.
The objective function of the first step to obtain the binary embedding $\mathbf{Y}$ is given as
\begin{align}
\min_{\mathbf{Z}, \mathbf{R}, \mathbf{Y}}~&~\operatorname{trace}(\mathbf{Z}^T\mathbf{M}\mathbf{Z}) + \eta \|\mathbf{Y} - \mathbf{Z}\mathbf{R}\|_F^2 \\
\operatorname{s.t.}~&~ \mathbf{Y} \in \{1, -1\}^{N \times M }, \mathbf{R}^T\mathbf{R} = \mathbf{I}.
\end{align}
Here $\mathbf{Z}$ is a nonlinear embedding, similar to locally linear embedding
and $\mathbf{M}$ is a sparse matrix, $\mathbf{M} = (\mathbf{I} - \mathbf{W})^T(\mathbf{I} - \mathbf{W})$.
$\mathbf{W}$ is the locally linear reconstruction weight matrix, which is computed
by solving the following optimization problem for each database item,
\begin{align}
\min_{\mathbf{w}_n} ~&~ \lambda \|\mathbf{s}_n^T\mathbf{w}_n\|_1 + \frac{1}{2} \|\mathbf{x}_n - \sum_{j \in \mathcal{N}(\mathbf{x}_n)} w_{ij}\mathbf{x}_n\|_2^2 \\
\operatorname{s.t.} ~&~ \mathbf{w}_n^T\mathbf{1} = 1,
\end{align}
where $\mathbf{w}_n = [w_{n1}, w_{n2}, \cdots, w_{nn}]^T$,
and $w_{nj} = 0$ if $j \notin \mathcal{N}(\mathbf{x}_n)$.
$\mathbf{s}_n = [s_{n1}, s_{n2}, \cdots, s_{nn}]^T$ is a vector
and $s_{nj} = \frac{\|\mathbf{x}_n - \mathbf{x}_j\|_2}{\sum_{t \in \mathcal{N}(\mathbf{x}_n)} \|\mathbf{x}_n - \mathbf{x}_t\|_2}$.

Out-of-sample extension computes the binary embedding of a query $\mathbf{q}$
as $\mathbf{y}_q = \operatorname{sign} (\mathbf{Y}^T\mathbf{w}_q)$.
Here $\mathbf{w}_q$ is a locally linear reconstruction weight,
and computed similarly to the above optimization problem.
Differently, $\mathbf{Y}$ and $\mathbf{w}_q$
correspond to the cluster centers, computed using $k$-means,
of the database $\mathbf{X}$.

\subsection{Beyond Hamming Distances in the Coding Space}
This section reviews the algorithms
focusing on designing effective distance measures given the binary codes
and possibly the hash functions.
The summary is given in Table~\ref{tab:lth:summaryofbeyondHammingDistance}.

\begin{table*}[t]
\centering
\caption{A summary of algorithms beyond Hamming distances in the coding space.}
\label{tab:lth:summaryofbeyondHammingDistance}
\begin{tabular}{|l||c||c|c|c|}
\hline
method & input similarity & distance measure\\
\hline
\hline
Manhattan hashing~\cite{KongLG12} & E & MD \\
Asymmetric distance I ~\cite{KongLG12} & E & AED, SED \\
Asymmetric distance II ~\cite{KongLG12} & E & LB  \\
asymmetric Hamming embedding~\cite{JainJG11} & E & LB \\
\hline
\end{tabular}
\end{table*}

\subsubsection{Manhattan distance}
When assigning multiple bits into a projection direction,
the Hamming distance breaks the neighborhood structure,
thus the points with smaller Hamming distance
along the projection direction
might have large Euclidean distance along the projection direction.
Manhattan hashing~\cite{KongLG12} introduces a scheme to address this issue,
the Hamming codes along the projection direction are in turn
(e.g., from the left to the right) transformed integers,
and the difference of the integers is used to replace the Hamming distance.
The aggregation of the differences along all the projection directions
is used as the distance of the hash codes.

\subsubsection{Asymmetric distance}
Let the compound hash function consist of
$K$ hash functions $\{h_k(\mathbf{x}) = b_k(g_k(\mathbf{x}))\}$,
where $g_k()$ is a real-valued embedding function
and $b_k()$ is a binarization function.
Asymmetric distance~\cite{GordoPGL14}presents two schemes.
The first one (Asymmetric distance I) is based on the expectation
$\bar{g}_{kb} = \operatorname{E}(g_k(\mathbf{x})| h_k(\mathbf{x})= b_k(g_k(\mathbf{x})) = b)$,
where $b=0$ and $b=1$.
When performing an online search,
a distance lookup table is precomputed:
\begin{align}
\{ & d_e(g_1(\mathbf{q}), \bar{g}_{10}),
d_e(g_1(\mathbf{q}), \bar{g}_{11}),
d_e(g_2(\mathbf{q}), \bar{g}_{20}),\nonumber \\
& d_e(g_2(\mathbf{q}), \bar{g}_{21}),
\cdots,
d_e(g_K(\mathbf{q}), \bar{g}_{K0}),
d_e(g_K(\mathbf{q}), \bar{g}_{K1}),
\end{align}
where $d_e(\cdot, \cdot)$ is an Euclidean distance operation.
Then the distance is computed
as $d_{ah}(\mathbf{q}, \mathbf{x})
= \sum_{k=1}^K d_e(g_k(\mathbf{q}), \bar{g}_{kh_k(\mathbf{x})})$,
which can be speeded up
e.g.,
by grouping the hash functions in blocks of $8$ bits
and have one $256$-dimensional look-up table per block
(rather than one $2$-dimensional look-up table per hash function.)
This reduces the number of summations
as well as the number of lookup operations.

The second scheme (Asymmetric distance II)  is
under the assumption that $b_k(g_k(\mathbf{x})) = \delta[g_k(\mathbf{x}) > t_k]$,
and computes the distance lower bound (similar way also adopted in asymmetric Hamming embedding~\cite{JainJG11}) over the $k$-th hash function,
\begin{equation}
d(g_k(\mathbf{q}), b_k(g_k(\mathbf{x})))
= \left\{ \begin{array}{l l}
     |g_k(\mathbf{q})| & \quad \text{if $h_k(\mathbf{x}) \neq h_k(\mathbf{q})$}\\
     0 & \quad \text{otherwise.}
   \end{array} \right.
\end{equation}
Similar to the first one,
the distance is computed
as $d_{ah}(\mathbf{q}, \mathbf{x})
= \sum_{k=1}^K d(g_k(\mathbf{q}), b_k(g_k(\mathbf{x})))$.
Similar hash function grouping scheme is used to speed up the search efficiency.


\subsubsection{Query sensitive hash code ranking}
Query sensitive hash code ranking~\cite{ZhangZS12} presented a similar asymmetric scheme
for $R$-neighbor search.
This method uses the PCA projection $\mathbf{W}$
to formulate the hash functions $\operatorname{sign}(\mathbf{W}^T\mathbf{x})= \operatorname{sign}(\mathbf{z})$.
The similarity along the $k$ projection is computed
as
\begin{align}
s_k(q_k, y_k, R) = \frac{P(z_k y_k > 0, |q_k - z_k | \leqslant R)}{P(|q_k - z_k | \leqslant R))},
\end{align}
which intuitively means that
the fraction of the points that lie in the range $|q_k - z_k | \leqslant R$
and are mapped to $y_k$
over the points that lie in the range $|q_k - z_k | \leqslant R$.
The similarity is computed
with the assumption that $p(z_k)$ is a Gaussian distribution.
The whole similarity is then
computed as $\prod_{k=1}^K s_k(q_k, y_k, R)$,
equivalently $\sum_{k=1}^K \log s_k(q_k, y_k, R)$.
The lookup table is also used to speed up the distance computation.

\subsubsection{Bit reconfiguration}
The goal of bits reconfiguration~\cite{MuCLCY12}
is to learn a good distance measure
over the hash codes precomputed
from a pool of hash functions.
Given the hash codes $\{\mathbf{y}_n\}_{n=1}^{N}$ with length $M$,
the similar pairs $\mathcal{M} = \{(i, j)\}$
and the dissimilar pairs $\mathcal{C} = \{(i, j)\}$,
compute the difference matrix $\mathbf{D}_m$ ($\mathbf{D}_c$)
over $\mathcal{M}$ ($\mathcal{C}$)
each column of which corresponds to $\mathbf{y}_i - \mathbf{y}_j$,
$(i, j) \in \mathcal{M}$ ($(i, j) \in \mathcal{C}$).
The formulation is given as the following maximization problem,
\begin{align}
\max_{\mathbf{W}}~&~\frac{1}{n_c}\operatorname{trace}(\mathbf{W}^T\mathbf{D}_c\mathbf{D}_c^T\mathbf{W})
- \frac{1}{n_m} \operatorname{trace}(\mathbf{W}^T\mathbf{D}_m\mathbf{D}_m^T\mathbf{W}) \nonumber \\
&~+ \frac{\eta}{n_s}\operatorname{trace}(\mathbf{W}^T\mathbf{Y}_s\mathbf{Y}_s^T\mathbf{W})
-\eta \operatorname{trace}(\mathbf{W}^T\boldsymbol{\mu}\boldsymbol{\mu}^T\mathbf{W}),
\end{align}
where $\mathbf{W}$ is a projection matrix of size $b \times t$.
The first term aims to maximize the differences between dissimilar pairs,
and the second term aims to minimize the differences between similar pairs.
The last two terms are maximized
so that the bit distribution is balanced,
which is derived
by maximizing $\operatorname{E}[\|\mathbf{W}^T(\mathbf{y} - \boldsymbol{\mu})\|_2^2]$,
where $\boldsymbol{\mu}$ represents the mean of the hash vectors,
and $\mathbf{Y}_s$ is a subset of input hash vectors with cardinality $n_s$.
\cite{MuCLCY12} furthermore refines the hash vectors
using the idea
of supervised locality-preserving method
based on graph Laplacian.

\section{Learning to Hash: Quantization}
\label{sec:LTH2}

\begin{table*}[t]
\centering
\caption{A summary of quantization algorithms.
sim. = similarity.
dist. = distance.}
\label{tab:lth:summaryofquantizatioalgorithms}
\begin{tabular}{|l||c||c|c|c|}
\hline
method & input sim. & hash function & dist. measure & optimization criteria\\
\hline
\hline
transform coding~\cite{Brandt10} & E & OQ & AED, SED & BA \\
double-bit quantization~\cite{KongL12b} & E & OQ & HD & $3$ partitions \\
\hline
iterative quantization~\cite{GongL11, GongLGP13} & E & LI & HD & QE \\
isotropic hashing~\cite{KongL12a} & E & LI & HD & EV \\
harmonious hashing~\cite{XuBLCHC13} & E & LI & HD & QE + EV \\
Angular quantization~\cite{GongKVL12} & CS & LI & NHA & MCS \\
\hline
product quantization~\cite{JegouDS11} & E & QU & (A)ED & QE \\
Cartesian $k$-means~\cite{NorouziF13} & E & QU & (A)ED & QE \\
composite quantization~\cite{ZhangDW14} & E & QU & (A)ED & QE \\
\hline
\end{tabular}
\end{table*}

This section focuses on the algorithms
that are based on quantization.
The representative algorithms are
summarized in~Table~\ref{tab:lth:summaryofquantizatioalgorithms}.

\subsection{$1$D Quantization}
This section reviews the hashing algorithms
that focuses on how to do the quantization along a projection direction
(partitioning the projection values of the reference data items along the direction
into multiple parts).

\subsubsection{Transform coding}
Similar to spectral hashing,
transform coding~\cite{Brandt10}
first transforms the data using PCA
and then assigns several bits to each principal direction.
Different from spectral hashing that uses Laplacian eigenvalues computed along each direction
to select Laplacian eigenfunctions to form hash functions,
transform coding
first adopts bit allocation
to determine which principal direction is used
and how many bits are assigned to such a direction.

The bit allocation algorithm is given as follows in Algorithm~\ref{alg:transformcoding}.
To form the hash function,
each selected principal direction $i$ is quantized
into $2^{m_i}$ clusters with the centers as $\{c_{i1}, c_{i2}, \cdots, c_{i2^{m_i}}\}$,
where each center is represented by a binary code of length $m_i$.
Encoding an item
consists of PCA projection
followed by quantization of the components.
the hash function can be formulated.
The distance between a query item and the hash code
is evaluated
as the aggregation
of the distance between the centers of the query and the database item
along each selected principal direction,
or the aggregation
of the distance between the center of the database item and the projection of the query of the corresponding principal direction
along all the selected principal direction.

\floatname{algorithm}{\small Algorithm}
\algsetup{
   linenosize=\small,
   linenodelimiter=.
}
\begin{algorithm}[t]
\footnotesize
\caption{\footnotesize Distribute $M$ bits into the principal directions}
\label{alg:iteratedlocalsearch}
\begin{algorithmic}[1]
\STATE Initialization: $e_i \leftarrow \log_2\sigma_i$, $m_i \leftarrow 0$.
\FOR{$j=1$ to $b$}
  \STATE $i \leftarrow \arg\max e_i$.
  \STATE $m_i \leftarrow m_i +1$.
  \STATE $e_i \leftarrow e_i - 1$.
\ENDFOR
\end{algorithmic}
\label{alg:transformcoding}
\end{algorithm}

\subsubsection{Double-bit quantization}
The double-bit quantization-based hashing algorithm~\cite{KongL12b}
distributes two bits into each projection direction
instead of one bit in ITQ or
hierarchical hashing~\cite{LiuWKC11}.
Unlike transform coding quantizing the points into $2^b$ clusters
along each direction,
double-bit quantization conducts $3$-cluster quantization,
and then assigns $01$, $00$, and $11$
to each cluster so that
the Hamming distance between the points belonging to neighboring clusters
is $1$,
and the Hamming distance between the points
not belonging to neighboring clusters
is $2$.


Local digit coding~\cite{KoudasOST04}
represents each dimension of a point
by a single bit,
which is set to $1$ if the value of the dimension
it corresponds to
is larger than a threshold
(derived from the mean of the corresponding data points),
and $0$ otherwise.

\subsection{Hypercubic Quantization}
Hypercubic quantization refers to
a category of algorithms
that quantize a data item
to a vertex in a hypercubic,
i.e., a vector belonging to $\{[y_1, y_2, \cdots, y_M]|y_m \in \{-1, 1\}\}$.

\subsubsection{Iterative quantization}
Iterative quantization~\cite{GongL11, GongLGP13}
aims to find the hash codes such that
the difference between the hash codes
and the data items,
by viewing each bit as the quantization value
along the corresponding dimension,
is minimized.
It consists of two steps:
(1) reduce the dimension using PCA to $M$ dimensions,
$\mathbf{v} = \mathbf{P}^T\mathbf{x}$,
where $\mathbf{P}$ is a matrix of size $d\times M$ ($M \leqslant d$)
computed using PCA,
and (2)
find the hash codes as well as an optimal rotation $\mathbf{R}$,
by solving the following optimization problem,
\begin{align}
\min\|\mathbf{Y} - \mathbf{R}^T\mathbf{V}\|_F^2, \label{eqn:ObjectiveFunctionForITQ}
\end{align}
where $\mathbf{V} = [\mathbf{v}_1 \mathbf{v}_2\cdots\mathbf{v}_N]$
and $\mathbf{Y} = [\mathbf{y}_1 \mathbf{y}_2\cdots \mathbf{y}_N]$.

The problem is solved via alternative optimization.
There are two alternative steps.
Fixing $\mathbf{R}$,
$\mathbf{Y} = \operatorname{sign}(\mathbf{R}^T\mathbf{V})$.
Fixing $\mathbf{B}$,
the problem becomes the classic orthogonal Procrustes problem,
and the solution is $\mathbf{R} = \hat{\mathbf{S}} \mathbf{S}^T$,
where $\mathbf{S}$ and $\hat{\mathbf{S}}$
is obtained
from the SVD of $\mathbf{Y}\mathbf{V}^T$,
$\mathbf{Y}\mathbf{V}^T = \mathbf{S}\boldsymbol{\Lambda}\hat{\mathbf{S}}^T$.

We present an integrated objective function
that is able to explain
the necessity of the first step.
Let $\bar{\mathbf{y}}$ be a $d$-dimensional vector,
which is a concatenated vector from $\mathbf{y}$ and an all-zero subvector:
$\bar{\mathbf{y}} = [\mathbf{y}^T 0... 0]^T$.
The integrated objective function is written as follows:
\begin{align}
\min\|\bar{\mathbf{Y}} - \bar{\mathbf{R}}^T\mathbf{X}\|_F^2,\label{eqn:IntegratedObjectiveFunctionForITQ}
\end{align}
where $\bar{\mathbf{Y}} = [\bar{\mathbf{y}}_1 \bar{\mathbf{y}}_2 \cdots \bar{\mathbf{y}}_N]$
$\mathbf{X} = [\mathbf{x}_1 \mathbf{x}_2 \cdots \mathbf{x}_N]$,
and $\bar{\mathbf{R}}$ is a rotation matrix.

Let $\bar{\mathbf{P}}$ be the projection matrix of $d \times d$,
computed using PCA,
$\bar{\mathbf{P}} = [\mathbf{P} \mathbf{P}_{-}]$.
It can be seen that,
the solutions for $\mathbf{y}$ of the two problems
in~\ref{eqn:IntegratedObjectiveFunctionForITQ} and~\ref{eqn:ObjectiveFunctionForITQ}
are the same,
if $\bar{\mathbf{R}} = \bar{\mathbf{P}}\operatorname{Diag}(\mathbf{R}, \mathbf{I})$.

%
%
%
%

\subsubsection{Isotropic hashing}
The idea of isotropic hashing~\cite{KongL12a}
is to rotate the space
so that the variance along each dimension is the same.
It consists of three steps:
(1) reduce the dimension using PCA to $M$ dimensions,
$\mathbf{v} = \mathbf{P}^T\mathbf{x}$,
where $\mathbf{P}$ is a matrix of size $d\times M$ ($M \leqslant d$)
computed using PCA,
and (2)
find an optimal rotation $\mathbf{R}$,
so that $\mathbf{R}^T\mathbf{V}\mathbf{V}^T\mathbf{R} = \boldsymbol{\Sigma}$
becomes a matrix with equal diagonal values,
i.e.,
$[\boldsymbol{\Sigma}]_{11} = [\boldsymbol{\Sigma}]_{22} = \cdots = [\boldsymbol{\Sigma}]_{MM}$.

Let $\sigma = \frac{1}{M}\operatorname{Trace}{\mathbf{V}\mathbf{V}^T}$.
The isotropic hashing algorithm then aims to find
an rotation matrix,
by solving the following problem:
\begin{align}
\|\mathbf{R}^T\mathbf{V}\mathbf{V}^T\mathbf{R} - \mathbf{Z}\|_F = 0,
\end{align}
where $\mathbf{Z}$ is a matrix with all the diagonal entries equal to $\sigma$.
The problem can be solved by two algorithms: lift and projection and gradient flow.

The goal of making the variances along the $M$ directions same
is to make the bits in the hash codes equally contributed to the distance evaluation.
In the case that
the data items satisfy the isotropic Gaussian distribution,
the solution from isotropic hashing is equivalent to iterative quantization.

Similar to generalized iterative quantization,
the PCA preprocess in isotropic hashing is also interpretable:
finding a global rotation matrix $\bar{\mathbf{R}}$ such that
the first $M$ diagonal entries of $\bar{\boldsymbol{\Sigma}}\bar{\mathbf{R}}^T\mathbf{X}\mathbf{X}^T\bar{\mathbf{R}}$
are equal,
and their sum is as large as possible,
which is formally written as follows,
\begin{align}
\max ~&~ \sum_{m=1}^M [\boldsymbol{\Sigma}]_{mm} \\
\operatorname{s.t.}
~&~ [\boldsymbol{\Sigma}] = \sigma, m=1,\cdots,M \\
~&~ \mathbf{R}^T\mathbf{R} = \mathbf{I}.
\end{align}

%
%
%

\subsubsection{Harmonious hashing}
Harmonious hashing~\cite{XuBLCHC13}
can be viewed as a combination of ITQ and Isotropic hashing.
The formulation is given as follows,
\begin{align}
\min_{\mathbf{Y}, \mathbf{R}}~&~\|\mathbf{Y} - \mathbf{R}^T\mathbf{V}\|_F^2 \\
\operatorname{s.t.}~&~\mathbf{Y}\mathbf{Y}^T=\sigma \mathbf{I} \\
~&~ \mathbf{R}^T\mathbf{R} = \mathbf{I}.
\end{align}

It is different from ITQ
in that the formulation does not require $\mathbf{Y}$ to be a binary matrix.
An iterative algorithm is presented to
optimize the above problem.
Fixing $\mathbf{R}$,
let $\mathbf{R}^T\mathbf{V} = \mathbf{\mathbf{U} \boldsymbol{\Lambda}\mathbf{V}^T}$,
then $\mathbf{Y} = \sigma^{1/2} \mathbf{U} \mathbf{V}^T$.
Fixing $\mathbf{\mathbf{Y}}$,
$\mathbf{R} = \hat{\mathbf{S}} \mathbf{S}^T$,
where $\mathbf{S}$ and $\hat{\mathbf{S}}$
is obtained
from the SVD of $\mathbf{Y}\mathbf{V}^T$,
$\mathbf{Y}\mathbf{V}^T = \mathbf{S}\boldsymbol{\Lambda}\hat{\mathbf{S}}^T$.
Finally, $\mathbf{Y}$ is cut at zero,
attaining binary codes.


\subsubsection{Angular quantization}
Angular quantization~\cite{GongKVL12}
addresses the ANN search problem
under the cosine similarity.
The basic idea
is to use the nearest vertex
from the vertices
of the binary hypercube $\{0, 1\}^d$
to approximate the data vector $\mathbf{x}$,
$\arg\max_{\mathbf{y}} \frac{\mathbf{y}^T\mathbf{x}}{\|\mathbf{b}\|_2}$,
subject to $\mathbf{y} \in \{0, 1\}^d$,
which is shown to be solved in $O(d \log d)$ time,
and then to evaluate the similarity
$\frac{\mathbf{b}_x^T \mathbf{b}_q}{\|\mathbf{b}_q\|_2 \|\mathbf{b}_x\|_2}$ in the Hamming space.

The objective function
of finding the binary codes,
similar to iterative quantization~\cite{GongL11},
is formulated as below,
\begin{align}
\max_{\mathbf{R}, \{\mathbf{y}_n\}}~&~
\sum_{n = 1}^N\frac{\mathbf{y}_n^T}{ \|\mathbf{y}_n\|_2} \frac{\mathbf{R}^T\mathbf{x}_n}{\|\mathbf{R}^T\mathbf{x}_n\|_2} \\
\operatorname{s.t.}~&~\mathbf{y}_n \in \{0, 1\}^M, \\
~&~\mathbf{R}^T\mathbf{R} = \mathbf{I}_M.
\end{align}
Here $\mathbf{R}$ is a projection matrix of $d \times M$.
This is transformed to
an easily-solved problem
by discarding the denominator $\|\mathbf{R}^T\mathbf{x}_n\|_2$:
\begin{align}
\max_{\mathbf{R}, \{\mathbf{y}_n\}}~&~
\sum_{n = 1}^N\frac{\mathbf{y}_n^T}{ \|\mathbf{y}_n\|_2} \mathbf{R}^T\mathbf{x}_n\\
\operatorname{s.t.}~&~\mathbf{y}_n \in \{0, 1\}^M, \\
~&~\mathbf{R}^T\mathbf{R} = \mathbf{I}_M.
\end{align}
The above problem is solved
using alternative optimization.

\subsection{Cartesian Quantization}

\subsubsection{Product quantization}
The basic idea of product quantization~\cite{JegouDS11}
is to divide the feature space
into ($P$) disjoint subspaces,
thus the database is divided into $P$ sets,
each set consisting of $N$ subvectors
$\{\mathbf{x}_{p1}, \cdots, \mathbf{x}_{pN}\}$,
and then to quantize each subspace separately
into ($K$) clusters.
Let $\{\mathbf{c}_{p1}, \mathbf{c}_{p2},
\cdots, \mathbf{c}_{pK}\}$ be the cluster centers
of the $p$ subspace,
each of which can be encoded
as a code of length $\log_2 K$.

A data item $\mathbf{x}_n$
is divided into $P$ subvectors
$\{\mathbf{x}_{pn}\}$,
and each subvector is assigned
to the nearest center $\mathbf{c}_{pk_{pn}}$
among the cluster centers
of the $p$th subspace.
Then the data item
$\mathbf{x}_n$ is represent by
$P$ subvectors $\{\mathbf{c}_{pk_{pn}}\}_{p=1}^P$,
thus represented
by a code of length $P\log_2 K$, $k_{1n}k_{2n} \cdots k_{Pn}$.
Product quantization can be viewed
as minimizing the following objective function,
\begin{align}
\min_{\mathbf{C}, \{\mathbf{b}_n\}} ~&~
\sum_{n=1}^N \|\mathbf{x}_n - \mathbf{C}\mathbf{b}_n\|_2^2.
\end{align}
Here $\mathbf{C}$ is a matrix of $d \times PK$
in the form of
\begin{align}
\operatorname{diag}(\mathbf{C}_1, \mathbf{C}_2, \cdots, \mathbf{C}_P)
=
\begin{bmatrix}
\mathbf{C}_1 & \mathbf{0} &  \cdots &  \mathbf{0}  \\[0.3em]
\mathbf{0} &  \mathbf{C}_2 &  \cdots &  \mathbf{0}  \\[0.3em]
\vdots  & \vdots  & \ddots & \vdots   \\[0.3em]
\mathbf{0} &  \mathbf{0} &  \cdots &  \mathbf{C}_P   \\
\end{bmatrix},
\end{align}
where $\mathbf{C}_p = [\mathbf{c}_{p1} \mathbf{c}_{p2}\cdots \mathbf{c}_{pK}]$.
$\mathbf{b}_n$ is the composition vector,
and its subvector
$\mathbf{b}_{np}$
of length $K$
is an indicator vector
with only one entry being $1$
and all others being $0$,
showing which element is selected
from the $p$th dictionary for quantization.

Given a query vector $\mathbf{x}_t$,
the distance to a vector $\mathbf{x}_n$,
represented by a code $k_{1n}k_{2n} \cdots k_{Pn}$
can be evaluated
in symmetric and asymmetric ways.
The symmetric distance is computed
as follows.
First, the code of the query
$\mathbf{x}_t$ is computed
using the way similar to the database vector,
denoted by
$k_{1t}k_{2t} \cdots k_{Pt}$.
Second,
a distance table is computed.
The table consists of $PK$ distance entries,
$\{d_{pk} = \|\mathbf{c}_{pk_{pt}} - \mathbf{c}_{pk}\|_2^2 | p = 1, \cdots, P, k = 1, \cdots, K\}$.
Finally,
the distance of the query to the vector $\mathbf{x}_n$
is computed by looking up the distance table
and summing up $P$ distances,
$\sum_{p=1}^P d_{pk_{pn}}$.
The asymmetric distance does not encode the query vector,
directly computes the distance table that also includes $PK$ distance entries,
$\{d_{pk} = \|\mathbf{x}_{pt} - \mathbf{c}_{pk}\|_2^2 | p = 1, \cdots, P, k = 1, \cdots, K\}$,
and finally conducts the same step to the symmetric distance evaluation,
computing the distance
as $\sum_{p=1}^P d_{pk_{pn}}$.

Distance-encoded product quantization~\cite{HeoLY14}
extends product quantization
by encoding both the cluster index
and the distance between a point and its cluster center.
The way of encoding the cluster index is similar to that in product quantization.
The way
of encoding the distance between a point and its cluster center
is given as follows.
Given a set of points belonging to a cluster,
those points are partitioned (quantized)
according to the distances to the cluster center.

\subsubsection{Cartesian $k$-means}
Cartesian $k$-means~\cite{NorouziF13, GeHK013} extends product quantization
and introduces a rotation $\mathbf{R}$
into the objective function,
\begin{align}
\min_{\mathbf{R}, \mathbf{C}, \{\mathbf{b}_n\}} ~&~
\sum_{n=1}^N \|\mathbf{R}^T\mathbf{x}_n - \mathbf{C}\mathbf{b}_n\|_2^2.
\end{align}
The introduced rotation does not
affect the Euclidean distance
as the Euclidean distance is invariant to the rotation,
and helps to find an optimized subspace partition
for quantization.

The problem is solved
by an alternative optimization algorithm.
Each iteration alternatively solves
$\mathbf{C}$,
$\{\mathbf{b}_n\}$,
and $\mathbf{R}$.
Fixing $\mathbf{R}$,
$\mathbf{C}$ and $\{\mathbf{b}_n\}$
are solved
using the same way as the one in product quantization
but with fewer iterations
and the necessity of reaching the converged solution.
Fixing $\mathbf{C}$ and $\{\mathbf{b}_n\}$,
the problem of optimizing $\mathbf{R}$
is the classic orthogonal Procrustes problem,
also occurring in iterative quantization.

The database vector $\mathbf{x}_n$ with Cartesian $k$-means
is represented by
$P$ subvectors $\{\mathbf{c}_{pk_{pn}}\}_{p=1}^P$,
thus encoded as $k_{1n}k_{2n} \cdots k_{Pn}$,
with a rotation matrix $\mathbf{R}$
for all database vector
(thus the rotation matrix does not increase the code length).
Given a query vector $\mathbf{x}_t$,
it is first rotated
as $\mathbf{R}^T\mathbf{x}_t$.
Then the distance is computed
using the same way to that in production quantization.
As rotating the query vector is only done once
for a query,
its computation cost for a large database
is negligible
compared with the cost of computing the approximate distances
with a large amount of database vectors.

Locally optimized product quantization~\cite{KalantidisA14}
applies Cartesian $k$-means to the search algorithm
with the inverted index,
where there is a quantizer for each inverted list.

\subsubsection{Composite quantization}
The basic ideas of composite quantization~\cite{ZhangDW14}
consist of (1) approximating the database vector
$\mathbf{x}_n$
using $P$ vectors with the same dimension $d$,
$\mathbf{c}_{1k_{1n}}, \mathbf{c}_{1k_{2n}}, \cdots, \mathbf{c}_{1k_{Pn}} $,
each selected from $K$ elements among one of $P$ source dictionaries $\{\mathcal{C}_1, \mathcal{C}_2, \cdots, \mathcal{C}_P\}$,
respectively,
(2) making the summation of the inner products of all
pairs of elements that are used to approximate the vector
but from different dictionaries,
$\sum_{i=1}^P\sum_{j=1, \neq i}^P \mathbf{c}_{ik_{in}} \mathbf{c}_{jk_{jn}}$,
be constant.

The problem is formulated as
\begin{align}\small
\min_{\{\mathbf{C}_p\},\{\mathbf{b}_n\}, \epsilon } ~&~\sum\nolimits_{n=1}^N\|\mathbf{x}_n -  [\mathbf{C}_1 \mathbf{C}_2 \cdots \mathbf{C}_P]\mathbf{b}_{n}\|_2^2 \label{eqn:originalobjectivefunction} \\
\operatorname{s.t.}
~&~ \sum\nolimits_{i=1}^P \sum\nolimits_{j=1, j\neq i}^P  \mathbf{b}_{ni}^T \mathbf{C}_{i}^T\mathbf{C}_{j}\mathbf{b}_{nj} = \epsilon \nonumber \\
~&~ \mathbf{b}_n  = [\mathbf{b}_{n1}^T\mathbf{b}_{n2}^T\cdots\mathbf{b}_{nP}^T]^T \nonumber \\
~&~ \mathbf{b}_{np} \in \{0, 1\}^K,~\|\mathbf{b}_{np}\|_1 = 1 \nonumber \\
~&~ n=1,2,\cdots,N, p=1,2, \cdots P.\nonumber
\end{align}
Here, ${\mathbf{C}_p}$ is a matrix of size $d \times K$,
and each column corresponds to an element
of the $p$th dictionary $\mathcal{C}_p$.

To get an easily optimization algorithm,
the objective function is transformed as
\begin{align}
&\phi({\{\mathbf{C}_p\}}, \{\mathbf{b}_n\}, \epsilon)  = \sum\nolimits_{n=1}^N\|\mathbf{x}_n -  \mathbf{C}\mathbf{b}_{n}\|_2^2 \nonumber\\
&+ \mu \sum\nolimits_{n=1}^N(\sum\nolimits_{i \neq j}^P \mathbf{b}_{ni}^T \mathbf{C}_{i}^T\mathbf{C}_{j}\mathbf{b}_{nj} - \epsilon)^2,
\label{eqn:objectivefunction}
\end{align}
where $\mu$ is the penalty parameter,
$\mathbf{C} = [\mathbf{C}_1 \mathbf{C}_2 \cdots \mathbf{C}_P]$
and $\sum\nolimits_{i \neq j}^P = \sum_{i=1}^P\sum_{j=1, j\neq i}^P$.
The transformed problem is solved
by alternative optimization.

The idea of using the summation
of several dictionary items
as an approximation of a data item
has already been studied
in the signal processing area,
known as multi-stage vector quantization,
residual quantization,
or more generally structured vector quantization~\cite{GrayN98},
and recently re-developed
for similarity search under the Euclidean distance~\cite{BabenkoK14, WangWSXSL14}
and inner product~\cite{DuW14}.

\section{Learning to Hash: Other Topics}
\label{sec:LTH3}
\subsection{Multi-Table Hashing}
\subsubsection{Complementary hashing}
The purpose of complementary hashing~\cite{XuWLZLY11}
is to learn multiple hash tables
such that nearest neighbors
have a large probability
to appear in the same bucket at least in one hash table.
The algorithm learns the hashing functions
for the multiple hash tables
in a sequential way.
The compound hash function for the first table
is learnt
by solving the same problem in~\cite{WangKC10a},
as formulated below
\begin{align}
\operatorname{trace}[\mathbf{W}^T\mathbf{X}_l\mathbf{S}\mathbf{X}_l^T\mathbf{W}]
+\eta \operatorname{trace}[\mathbf{W}^T\mathbf{X}\mathbf{X}^T\mathbf{W}],
\end{align}
where $s_{ij}$ is initialized
as $K(a_{ij} - \alpha)$,
$a_{ij}$ is the similarity between
$\mathbf{x}_i$ and $\mathbf{x}_j$
and $\alpha$ is a super-constant.

To compute the second compound hash functions,
the same objective function is optimized
but with different matrix $\mathbf{S}$:
\begin{equation}
s^{t}_{ij} =
\left\{ \begin{array}{l l}
     0 & \quad b^a_{ij} = b^{(t-1)}_{ij} \\
     \min(s_{ij}, f_{ij}) & \quad b^a_{ij} = 1, b^{(t-1)}_{ij} = -1 \\
     -\min(-s_{ij}, f_{ij}) & \quad b^a_{ij} = -1, b^{(t-1)}_{ij} = 1 \\
   \end{array} \right.
\end{equation}
where $f_{ij} = (a_{ij} - \alpha) (\frac{1}{4} d^{(t-1)}_h(\mathbf{x}_i, \mathbf{x}_j) - \beta)$,
$\beta$ is a super-constant,
and $
b^{(t-1)_{ij}} =
1
-2 \operatorname{sign}[\frac{1}{4}d^{(t-1)}_h(\mathbf{x}_i, \mathbf{x}_j) - \beta] $.
Some tricks are also given to scale up the problem
to large scale databases.

\subsubsection{Reciprocal hash tables}
The reciprocal hash tables~\cite{LiuHL13} extends
complementary hashing
by building a graph over a pool $B$ hash functions
(with the output being a binary value)
and searching the best hash functions over such a graph for building a hashing table,
updating the graph weight using a boosting-style algorithm
and finding the subsequent hash tables.
The vertex in the graph
corresponds to a hash function
and is associated with a weight showing the degree
that similar pairs are mapped to the same binary value
and dissimilar pairs are mapped to different binary values.
The weight over the edge connecting two hash functions
reflects the independence between two hash functions
the weight is higher if
the difference of the distributions of the binary values $\{-1, 1\}$ computed from the two hash functions
is larger.
\cite{LiuHLC13}
shows how to formulate the hash bit selection problem
into a quadric program,
which is derived from organizing the candidate bits in graph.

\subsection{Active and Online Hashing}
\subsubsection{Active hashing}
Active hashing~\cite{ZhenY13}
starts with a small set of pairs of points with labeling information
and actively selects the most informative labeled pairs
for hash function learning.
Given the sets of
labeled data $\mathcal{L}$,
unlabeled data $\mathcal{U}$,
and candidate data $\mathcal{C}$,
the algorithm first learns the compound hash function
$\mathbf{h} = \operatorname{sign}(\mathbf{W}^T\mathbf{x})$,
and then computes the data certainty score for each point in the candidate set,
$f(\mathbf{x}) = \|\mathbf{W}^T\mathbf{x}\|_2$,
which reflects the distance of a point to the hyperplane forming the hash functions.
Points with smaller the data certainty scores should be selected for further labeling.
On the other hand, the selected points should not be similar to each other.
To this end,
the problem of finding the most informative points is formulated
as the following,
\begin{align}
\min_{\mathbf{b}}~&~ \mathbf{b}^T\bar{\mathbf{f}} + \frac{\lambda}{M} \mathbf{b}^T\mathbf{K}\mathbf{b} \\
\operatorname{s.t.}~&~ \mathbf{b} \in \{0, 1\}^{\|\mathcal{C}\|} \\
~&~ \|\mathbf{b}^T\|_1 = M,
\end{align}
where $\mathbf{b}$ is an indicator vector
in which $b_i = 1$ when $\mathbf{x}_i$ is selected
and $b_i = 0$ when $\mathbf{x}_i$ is not selected,
$M$ is the number of points that need to be selected,
$\bar{\mathbf{f}}$ is a vector of the normalized certainty scores over the candidate set,
with each element $\bar{f}_i = \frac{f_i}{\max_{j=1}^{\|\mathcal{C}\|} \bar{f}_j}$,
$\mathbf{K}$ is the similarity matrix computed over $\mathcal{C}$,
and $\lambda$ is the trade-off parameter.

\subsubsection{Online hashing}
Online hashing~\cite{HuangYZ13} presents an algorithm
to learn the hash functions
when the similar/dissimilar pairs come
sequentially
rather than at the beginning,
all the similar/dissimilar pairs come together.
Smart hashing~\cite{YangHZL13} also addresses the problem
when the similar/dissimilar pairs come
sequentially.
Unlike the online hash algorithm
that updates all hash functions,
smart hashing only selects a small subset of hash functions
for relearning
for a fast response to newly-coming labeled pairs.

\subsection{Hashing for the Absolute Inner Product Similarity}
\subsubsection{Concomitant hashing}
Concomitant hashing~\cite{MuWC12} aims to find the points
with the smallest and largest absolute cosine similarity.
The approach is similar to concomitant LSH~\cite{EshghiR08}
and formulate a two-bit hash code
using a multi-set
$\{h_{min}(\mathbf{x}), h_{max}(\mathbf{x})\} = \{\arg\min_{k=1}^{2^K} \mathbf{w}_k^T\mathbf{x}, \arg\max_{k=1}^{2^K} \mathbf{w}_k^T\mathbf{x}\}$.
The two bits are unordered,
which is slightly different from concomitant LSH~\cite{EshghiR08}.
The collision probability is defined as
$\operatorname{Prob}[\{h_{min}(\mathbf{x}), h_{max}(\mathbf{x})\} =  \{h_{min}(\mathbf{y}), h_{max}(\mathbf{y})\}]$,
which is shown to be a monotonically increasing function
with respect to $|\mathbf{x}_1^T \mathbf{x}_2|$.
This, thus,
means that
the larger hamming distance, the smaller $|\mathbf{x}_1^T \mathbf{x}_2|$ (min-inner-product)
and
the smaller hamming distance, the larger $|\mathbf{x}_1^T \mathbf{x}_2|$ (max-inner-product).

\subsection{Matrix Hashing}
\subsubsection{Bilinear projection}
A bilinear projection algorithm is proposed in~\cite{GongKRL13}
to hash a matrix feature to short codes.
The (compound) hash function
is defined as
\begin{align}
\operatorname{vec}(\operatorname{sign}(\mathbf{R}_l^T\mathbf{X}\mathbf{R}_r)),
\end{align}
where $\mathbf{X}$
is a matrix of $d_l \times d_r$,
$\mathbf{R}_l$ of size $d_l \times d_l$  and $\mathbf{R}_r$
of size $d_r \times d_r$
are two random orthogonal matrices.
It is easy to show that
\begin{align}
\operatorname{vec}(\mathbf{R}_l^T\mathbf{X}\mathbf{R}_r)
= (\mathbf{R}_r^T \otimes \mathbf{R}_l^T) \operatorname{vec}(\mathbf{X})
= \mathbf{R}^T \operatorname{vec}(\mathbf{X}).
\end{align}

The objective is to minimize
the angle between a rotated feature
$\mathbf{R}^T \operatorname{vec}(\mathbf{X})$
and its binary encoding
$\operatorname{sign}(\mathbf{R}^T \operatorname{vec}(\mathbf{X}))
= \operatorname{vec}(\operatorname{sign}(\mathbf{R}_l^T\mathbf{X}\mathbf{R}_r))$.
The formulation is given as follows,
\begin{align}
\max_{\mathbf{R}_l, \mathbf{R}_r, \{\mathbf{B}_n\}}~&~
\sum_{n=1}^N \operatorname{trace}(\mathbf{B}_n \mathbf{R}_r^T \mathbf{X}_n^T \mathbf{R}_l)\\
\operatorname{s.t.}
~&~
\mathbf{B}_n \in \{-1, +1\}^{d_l \times d_r}\\
~&~ \mathbf{R}_l^T \mathbf{R}_l = \mathbf{I}\\
~&~ \mathbf{R}_r^T \mathbf{R}_r = \mathbf{I},
\end{align}
where $\mathbf{B}_n = \operatorname{sign}(\mathbf{R}_l^T\mathbf{X}_n\mathbf{R}_r)$.
The problem is optimized
by alternating
between $\{\mathbf{B}_n\}$,
$\mathbf{R}_l$
and
$\mathbf{R}_r$.
To reduce the code length,
the low-dimensional orthogonal matrices can be used:
$\mathbf{R}_l \in \mathbb{R}^{d_l \times c_l}$
and
$\mathbf{R}_r \in \mathbb{R}^{d_r \times c_r}$.

\subsection{Compact Sparse Coding}
Compact sparse coding~\cite{Cherian14},
the extension of the early work robust sparse coding~\cite{CherianMP12}
adopts sparse codes to represent
the database items:
the atom indices corresponding to nonzero codes are
used to build the inverted index,
and the nonzero coefficients
are used to
reconstruct the database items
and compute the approximate distances
between the query and the database items.

The sparse coding objective function,
with introducing the incoherence constraint
of the dictionary,
is given as follows,
\begin{align}
\min_{\mathbf{C}, \{\mathbf{z}_n\}_{n=1}^N}~&~\frac{1}{2} \sum_{i=1}^N\| \mathbf{x}_i - \sum_{j=1}^K z_{ij}\mathbf{c}_j \|_2^2 + \lambda \|\mathbf{z}_i\|_1 \\
\operatorname{s.t.} ~&~ \|\mathbf{C}^T_{\sim k}\mathbf{c}_k\|_{\infty} \leqslant \gamma; k=1,2, \cdots, n,
\end{align}
where $\mathbf{C}$ is the dictionary,
$\mathbf{C}^T_{\sim k}$ is the dictionary $\mathbf{C}$ with the $k$th atom removed,
$\{\mathbf{z}_n\}_{n=1}^N$ are the $N$ sparse codes.
$\|\mathbf{C}^T_{\sim k}\mathbf{c}_k\|_{\infty} \leqslant \gamma$ aims to control the dictionary coherence degree.

The support of $\mathbf{x}_n$ is defined as the indices
corresponding to nonzero coefficients in $\mathbf{z}_n$:
$\mathbf{b}_n = \delta.[\mathbf{z}_n \neq 0]$,
where $\delta.[]$ is an element-wise operation.
The introduced approach uses $\{\mathbf{b}_n\}_{n=1}^N$
to build the inverted indices, which is similar to min-hash,
and also uses the Jaccard similarity to get the search results.
Finally, the asymmetric distances between the query and the retrieved results using the Jaccard similarity,
$\|\mathbf{q} - \mathbf{B}\mathbf{z}_n\|_2$
are computed
for reranking.

\subsection{Fast Search in Hamming Space}

\subsubsection{Multi-index hashing}
The idea~\cite{NorouziPF12} is that
binary codes in the reference database
are indexed $M$ times
into $M$ different hash tables,
based on $M$ disjoint binary substrings.
Given a query binary code,
entries that fall close to the query
in at least one substring are considered neighbor candidates.
Specifically,
each code $\mathbf{y}$ is split
into $M$ disjoint subcodes
$\{\mathbf{y}^1, \cdots, \mathbf{y}^M\}$.
For each subcode, $\mathbf{y}^m$,
one hash table is built,
where each entry corresponds to a list
of indices of the binary code
whose $m$th subcodes is equal to the code associated with this entry.

To find $R$-neighbors of a query $\mathbf{q}$
with substrings $\{\mathbf{q}^m\}_{m=1}^M$,
the algorithm
search $m$th hash table
for entries that are within a Hamming distance $\lfloor \frac{R}{M} \rfloor$
of $\mathbf{q}^m$,
thereby retrieving a set of candidates, denoted by $\mathcal{N}_{m}(\mathbf{q})$
and thus a set of final candidates,
$\mathcal{N} = \cup_{m=1}^M \mathcal{N}_{m}(\mathbf{q})$.
Lastly,
the algorithm computes the Hamming distance between $\mathbf{q}$
and each candidate,
retaining only those codes that are true $R$-neighbors of $\mathbf{q}$.
\cite{NorouziPF12} also discussed how to choose the optimal number $M$ of substrings.

\subsubsection{FLANN}
\cite{MujaL12} extends the FLANN algorithm~\cite{MujaL09}
that is initially designed for ANN search
over real-value vectors
to search over binary vectors.
The key idea is to build multiple hierarchical cluster trees to organize the binary vectors
and search for the nearest neighbors
simultaneously over the multiple trees.

The tree building process starts with all the points
and divides them into $K$ clusters
with cluster centers randomly selected from the input points
and each point assigned to the center that is closest to the point.
The algorithm is repeated recursively
for each of the resulting clusters
until the number of points
in each cluster is below a certain threshold,
in which case that node becomes a leaf node.
The whole process is repeated several times,
yielding multiple trees.

The search process starts with
a single traverse of each of the trees,
during which the algorithm always picks the node
closest to the query point
and recursively explores it,
while adding the unexplored nodes to a priority queue.
When reaching the leaf node
all the points contained within are linearly searched.
After each of the trees has been explored once,
the search is continued by extracting from the priority queue
the closest node to the query point and resuming the tree traversal from there.
The search ends when the number of points examined exceeds a maximum limit.

%
%

\section{Discussions and Future Trends}

\subsection{Scalable Hash Function Learning}
The algorithms depending on the pairwise similarity,
such binary reconstructive embedding,
usually sample a small subset of pairs
to reduce the cost of learning hash functions.
It is shown that
the search accuracy is increased
with a high sampling rate,
but the training cost is greatly increased.
The algorithms even without relying pairwise similarity
are also shown to be slow and even infeasible
when handling very large data, e.g., $1B$ data items,
and usually learn hash functions over a small subset, e.g., $1M$ data items.
This poses a challenging request
to learn the hash function over larger datasets.

\subsection{Hash Code Computation Speedup}
Existing hashing algorithms rarely do not take
consideration of the cost of encoding a data item.
Such a cost during the query stage becomes significant
in the case
that only a small number of database items
or a small database
are compared to the query.
The search with
combining inverted index and compact codes
is such a case.
an recent work,
circulant binary embedding~\cite{YuKGC14},
formulates the projection matrix (the weights in the hash function)
using a circular matrix $\mathbf{R} = \operatorname{circ}(\mathbf{r})$.
The compound hash function is formulated
is given as $\mathbf{h}(\mathbf{x}) = \operatorname{sign}(\mathbf{R}^T\mathbf{x})$,
where the computation is accelerated
using fast Fourier transformation
with the time cost reduced from $O(d^2)$ to $d \log d$.
It expects more research study
to speed up the hash code computation for other hashing algorithms,
such as composite quantization.

\subsection{Distance Table Computation Speedup}
Product quantization and its variants
need to precompute the distance table
between the query and the elements of the dictionaries.
Existing algorithms claim
the cost of distance table computation is negligible.
However in practice, the cost becomes bigger
when using the codes computed from quantization
to rank the candidates retrieved from inverted index.
This is a research direction that will attract research interests.

\subsection{Multiple and Cross Modality Hashing}
One important characteristic of big data is the variety of data types and data sources. This
is particularly true to multimedia data, where various media types (e.g., video, image, audio
and hypertext) can be described by many different low- and high-level features, and relevant
multimedia objects may come from different data sources contributed by different users and
organizations.
This raises a research direction,
performing joint-modality hashing learning
by exploiting the relation among multiple modalities,
for supporting some special applications, such as
cross-model search.
This topic is attracting a lot of research efforts,
such as
collaborative hashing~\cite{LiuHDL14},
and cross-media hashing~\cite{SongYHSL13, SongYYHS13, ZhuHSZ13}.

\section{Conclusion}
\label{sec:con}
In this paper,
we review two categories of hashing algorithm
developed for similarity search:
locality sensitive hashing
and learning to hash
and show how they are designed
to conduct similarity search.
We also point out the future trends of hashing for similarity search.

{\small
\bibliographystyle{ieee}
\bibliography{hash}
}

\end{document}